\documentclass[fleqn,usenatbib]{mnras}

\usepackage{newtxtext,newtxmath}

\usepackage[T1]{fontenc}
\usepackage{ae,aecompl}
\usepackage{pdflscape}

\usepackage{graphicx}	
\usepackage{amsmath}	

\renewcommand{\deg}{\hbox{$^\circ$}}
\renewcommand{\arcmin}{\hbox{$^\prime$}}
\renewcommand{\arcsec}{\hbox{$^{\prime\prime}$}}

\defcitealias{williams2019lofar}{W19}
\defcitealias{kondapally2021}{Kondapally et al. 2021}
\defcitealias{nisbet2018}{Nisbet 2018}

\title[LOFAR machine learning classifier]{A machine learning classifier for LOFAR radio galaxy cross-matching techniques}

\author[Alegre, Sabater, Best et al.]{Lara Alegre,$^{1}$\thanks{E-mail: alegre@roe.ac.uk}
Jose Sabater,$^{1,2}$
Philip Best,$^{1}$
Rafa{\"e}l I.J. Mostert,$^{3,4}$
Wendy L. Williams,$^{4}$
\newauthor
G{\"u}lay G{\"u}rkan,$^{5}$
Martin J. Hardcastle,$^{6}$
Rohit Kondapally,$^{1}$
Tim W. Shimwell,$^{4,3}$
\newauthor
Daniel J.B. Smith$^{6}$
\\
$^{1}$SUPA, Institute for Astronomy, University of Edinburgh, Royal Observatory, Blackford Hill, Edinburgh, EH9 3HJ, UK\\
$^{2}$UK Astronomy Technology Centre, Royal Observatory, Blackford Hill, Edinburgh, EH9 3HJ, UK\\
$^{3}$Leiden Observatory, Leiden University, PO Box 9513, NL-2300 RA Leiden, The Netherlands\\
$^{4}$ASTRON, the Netherlands Institute for Radio Astronomy, Oude Hoogeveensedijk 4, 7991 PD Dwingeloo, The Netherlands\\
$^{5}$Th\"uringer Landessternwarte Tautenburg (TLS), Sternwarte 5, D-07778 Tautenburg, Germany\\
$^{6}$Centre for Astrophysics Research, Department of Physics, Astronomy and Mathematics, University of Hertfordshire, College Lane, Hatfield AL10 9AB
}
\date{Accepted XXX. Received YYY; in original form ZZZ}

\pubyear{2022}

\begin{document}
\label{firstpage}
\pagerange{\pageref{firstpage}--\pageref{lastpage}}
\maketitle



\begin{abstract}

New-generation radio telescopes like LOFAR are conducting extensive sky surveys, detecting millions of sources. To maximise the scientific value of these surveys, radio source components must be properly associated into physical sources before being cross-matched with their optical/infrared counterparts. In this paper, we use machine learning to identify those radio sources for which either source association is required or statistical cross-matching to optical/infrared catalogues is unreliable. We train a binary classifier using manual annotations from the LOFAR Two-metre Sky Survey (LoTSS). We find that, compared to a classification model based on just the radio source parameters, the addition of features of the nearest-neighbour radio sources, the potential optical host galaxy, and the radio source composition in terms of Gaussian components, all improve model performance. Our best model, a gradient boosting classifier, achieves an accuracy of 95 per cent on a balanced dataset and 96 per cent on the whole (unbalanced) sample after optimising the classification threshold. Unsurprisingly, the classifier performs best on small, unresolved radio sources, reaching almost 99 per cent accuracy for sources smaller than 15 arcsec, but still achieves 70 per cent accuracy on resolved sources. It flags 68 per cent more sources than required as needing visual inspection, but this is still fewer than the manually-developed decision tree used in LoTSS, while also having a lower rate of wrongly accepted sources for statistical analysis. The results have an immediate practical application for cross-matching the next LoTSS data releases and can be generalised to other radio surveys.  

\end{abstract}




\section{Introduction}
\label{sec:introduction}

The number of detected sources and the complexity of the structures in astronomical images has increased dramatically in recent years, with high sensitivity telescopes surveying deeper but also wider areas of the sky. Radio astronomy has been at the forefront of this big data revolution, with telescopes like the LOw Frequency ARray \citep[LOFAR,][]{vanHaarlen2013lofar}, the Very Large Array, and the Australian Square Kilometre Array Pathfinder Telescope \citep[ASKAP,][]{Hotan2021}. These have been conducting wide radio continuum surveys, such as the LOFAR Two-meter Sky Survey \citep[LoTSS,][]{Shimwell2017,Shimwell2019lofar,Shimwell2022}, the VLA Sky Survey \citep[VLASS, ][]{Lacy2020VLASS} and the Rapid ASKAP Continuum Survey \citep[RACS,][]{Hale2021}, and the Evolutionary Map of the Universe \citep[EMU,][]{norris2011emu}, respectively. When completed, these surveys will have covered both hemispheres and discovered tens of millions of radio sources.
This brings radio astronomy into a revolutionary new era: large samples enable detailed statistical studies whilst probing the unexplored Universe at these wavelengths \citep[see][for a review]{Norris2017Review}. In addition to producing scientific results, these surveys are also developing technology in preparation for the upcoming Square Kilometer Array \citep[SKA,][]{dewdney2009square}, which will be the world's most powerful radio telescope. The SKA will generate massive amounts of data and is expected to detect billions of radio sources.

In order to extract the full scientific return from these surveys, it is essential to cross-match the objects detected at radio wavelengths to their counterparts at other wavelengths, particularly optical and near-infrared. This allows us to identify the host galaxies, classify the radio sources according to their morphology, black hole activity, and other characteristics, and derive basic physical properties such as redshifts, luminosities and stellar masses \citep[e.g.][]{Best2005, Smolcic2017VLACosmos, Duncan2019lofar, Gulay2022}. The cross-identification of radio galaxies with their optical (or infrared) counterparts is a complex process due to the extended and multi-component nature of many radio sources, as well as the mismatch in the angular resolution between the radio and optical surveys. Traditionally, it has relied mostly on statistical methods, visual analysis, or a combination of the two \citep[see][hereafter referred as \citetalias{williams2019lofar}, for a discussion]{williams2019lofar}. 

In early continuum radio surveys the sources detected were mainly bright active galactic nuclei (AGN); only a small proportion of these had counterparts in the all-sky optical imaging data available at that time, but the samples were small enough that dedicated deep optical imaging of individual sources could be coupled with visual analysis \citep[e.g.][]{Laing1983}. By the turn of the century, a statistical comparison of the Faint Images of the Radio Sky at Twenty centimeters survey \citep[FIRST,][]{becker1995} with the large-area optical imaging from the Sloan Digital Sky Survey \citep[SDSS,][]{York2000} provided optical identifications for around 30 per cent of the $\sim 10^5$ radio source host galaxies \citep{Ivezic2002}. Recent radio surveys have been revealing still fainter sources, including higher fractions of star forming galaxies (SFG) that begin to dominate over AGN at low flux densities. At the same time, deeper optical and near-infrared observations are now available over large sky areas, such as imaging from the Panoramic Survey Telescope and Rapid Response System (Pan-STARRS-1) survey \citep[][]{Chambers2016} or the Dark Energy Spectroscopic Instrument (DESI) Legacy survey \citep{Dey_legacy2019}, with even deeper and wider imaging expected in the coming years from the Large Survey of Space and Time \citep[LSST;][]{Ivezic2019lsst} and the Euclid Space Telescope surveys \citep[][]{Laureijs2011euclid}. These surveys increase both the fraction of radio sources with optical counterparts and the number of potentially confusing foreground or background sources. The simultaneous increase of possible matches and data volumes requires improvement in the current cross-matching techniques.

In LoTSS, the source density is already more than a factor of 10 times higher than in the existing widely-used large-area radio continuum surveys such as the National Radio Astronomy Observatory (NRAO) VLA Sky Survey \citep[NVSS, ][]{condon1998nrao}, the FIRST survey, the Sydney University Molonglo Sky Survey \citep[SUMSS, ][]{Bock1999summs}, and the Westerbork Northern Sky Survey \citep[WENSS, ][]{rengelink1997westerbork}. 
LoTSS detected more than 300,000 sources in its first data release, containing just the first 2 per cent of the survey \citep[LoTSS DR1;][]{Shimwell2019lofar}, and a second data release with almost 4.4 million sources covering 5634 deg$^2$, 27 per cent of the northern sky, has just been published \citep[LoTSS DR2; ][]{Shimwell2022}. 

In LoTSS DR1, the radio sources were cross-matched with optical and near-infrared surveys, Pan-STARRS1 DR1 \citep{Chambers2016} and the AllWISE catalogue \citep{Cutri2013AllWISE}, respectively, and an optical and/or near-infrared counterpart was identifiable for 73 per cent of the LoTSS sources \citepalias{williams2019lofar}. Compact sources, such as SFGs or compact AGNs, were cross-matched using the Likelihood Ratio technique \citep[LR; e.g.,][]{Richter1975LR,Ruiter1977LR, Sutherland1992, Ciliegi2003LR} which assesses the relative probability of a given optical source being a true counterpart against a randomly aligned optical object, based on source properties \citepalias[for LoTSS DR1, the LR assessment considered both the magnitude and colour of the potential host galaxy; see][]{nisbet2018,williams2019lofar}. This statistical method is reliable when the flux-weighted mean position of the radio emission is an accurate estimate of the location at which the radio source originates, and is therefore coincident with the optical emission.
However, more extended sources cannot yet be reliably handled through these statistical methods. Furthermore, for radio sources with emission that is extended and/or split into different radio components (e.g. double-lobed sources), source detection algorithms often fail to correctly group together the multiple radio components into a single source, generating independent entries in the radio catalogues. In other cases, the source finder can incorrectly group individual physical radio sources together into a single blended detection. Thus, radio catalogues are not always a true description of the physical sources, leading to further inaccuracies if statistical techniques are naively applied. In LoTSS DR1, these complex-structured, multi-component and blended sources were therefore visually cross-matched alongside manual component association or dissociation.

In order to discriminate between sources that require visual analysis and those that can be reliably cross-matched using the LR technique, \citetalias{williams2019lofar} designed a decision tree based on the properties of the radio sources and their cross-ID LR values. This decision tree selected nearly 30,000 sources for visual inspection, corresponding to around 10 per cent of the total LoTSS DR1 sample. This was a conservative selection process, and indeed post-analysis (i.e. the examination of which ones actually required visual inspection, explained in Sec.~\ref{sec:data}) shows that only just over half of these sources actually required to be inspected. LoTSS DR2 covers an area almost 15 times larger than DR1, with a higher fraction of counterparts expected due to the use of the (deeper) Legacy dataset for cross-matching. The large number of sources makes visual inspection very challenging for more than a small fraction of the sources; while the ultimate goal is to replace all visual analysis with automated techniques, a more practical and immediate step is to minimise the amount of unnecessary inspection.

Some progress has been made to improve the current statistical methods, for example by modifying the LR technique to tackle the blending problem \citep{Weston2018}, by replacing the LR by Bayesian approaches \citep{Fan2015, Mallinar2017, fan2020optimal}, or by applying Machine Learning (ML) techniques \citep[e.g.][]{Alger2018}.
Various efforts have also been made to improve the cross-matching process for the extended/multi-component radio sources, using a ridgeline approach \citep{barkus2022application} and deep learning techniques mainly based on Convolutional Neural Networks (CNN), for example to group radio source components \citep{Mostert2022} or to find the host galaxy in previously-selected sources with multiple radio components \citep{Alger2018}. 
CNNs have also been used to improve the source finding and identification \citep[e.g.][]{vafaei2019deepsource}, or for automatic source extraction and further morphology classification \citep[e.g.][]{wu2019Claran}. 
Deep learning has been particularly successful in automating radio galaxy morphology classification of (previously associated) multi-component sources using CNNs \citep[e.g. ][]{Lukic2018,lukic2019morphological, Aniyan2017, alhassan2018first}, using transfer learning \citep{tang2019transfer} and using clustering methods \citep{galvin2020cataloguing, mostert2021unveiling} combined, for example, with Haralick features  \citep{ntwaetsile2021rapid}.
However, deep learning models, which perform feature extraction from the images before classification, require a higher number of annotated examples to train, and are also more difficult to interpret and to adapt than simpler ML models. In addition, a variety of unforeseen limitations due to limited experimentation in radio astronomy can further introduce different biases. Some examples include issues related to the use of fixed-size data images \citep[][]{mostert2021unveiling} or even the image input file format \citep[][]{tang2019transfer}. 
Furthermore, none of these methods can yet perform reliable source association and fully cross-match extended and multi-component sources. To date, the full cross-matching of modern large radio surveys has been only achieved through citizen science projects \citep[e.g. Radio Galaxy Zoo (RGZ), ][]{Banfield2015B_RGZ} and extensive science team efforts 
\citepalias[e.g. LOFAR Galaxy Zoo (LGZ),][]{williams2019lofar, kondapally2021}.

In this work, we propose a Gradient Boosting Classifier (GBC) to identify which radio sources can be reliably cross-matched using the LR technique, or instead require visual inspection. We use supervised ML algorithms, which offer greater intuitive interpretation and are simpler to adjust and analyse than deep learning models. The model adopted is an ensemble of decision trees, and it was selected and optimised using Automated Machine Learning \citep[AutoML; see Appendix~\ref{app:ML} and][for a review]{he2021automl}. While individual decision trees have been used in radio galaxy classification in the past \citep[e.g.][]{proctor2016selection}, ensembles of decision trees have been proven to achieve better performance  \citep[][]{dietterich2000experimental}. Examples of the use of ensembles of decision trees in radio astronomy include the classification of blazars using multiwavelength data \citep[][]{Arsioli2020} and the estimation of physical properties of radio sources such as redshifts \citep[][]{luken2022estimating}.

We build a dataset based on LoTSS DR1, which provides more than 300,000 annotated examples, and select a set of relevant features, allowing the model to successfully classify unseen sources with an accuracy of 94.6 per cent and select the ones that can be cross-matched by LR with a precision of 96.3 per cent. This helps to limit the manual analysis to the most complex sources (extended sources, sources with multiple components or blended detections), which are those for which the LR method is not successful. 
The results of this study are already being incorporated, by helping to identify unrelated radio components, into the automatic component association of sources larger than 15 arcsec from LoTSS DR1 \citep[][]{Mostert2022}. Furthermore, the methods applied in LoTSS DR1 are directly transferable to other parts of the LoTSS survey since the techniques used for processing and cross-matching the next data releases are broadly similar. Therefore, our work has immediate practical benefit for deciding which sources require visual analysis in LoTSS DR2 (Hardcastle et al. in prep). 

The paper is organised as follows: in Section \ref{sec:data} we describe the LoTSS DR1 data and in Section \ref{sec:machine_learning} we explain how these data were used to create a dataset suitable for our ML classification problem. Section \ref{sec:experiments} refers to the experiments performed to select and optimise the model, including the specifications of the model adopted. The model performance and interpretation are explained in Section \ref{sec:model_performance_and_interpretation}. In Section \ref{sec:implications_for_DR1} we interpret the results of the model applied to the full LoTSS datasets, discussing the implications and comparing them against the methods currently used. The conclusions and a discussion of their significance for the next LoTSS data releases can be found in Section \ref{sec:conclusions}. 
\section{Data}
\label{sec:data}

The data used in this work consist of LoTSS DR1 \citep{Shimwell2019lofar}\footnote{\hyperlink{https://lofar-surveys.org}{https://lofar-surveys.org}} radio catalogues that were derived from the 58 mosaic images of DR1, which cover 424\,deg$^2$ over the Hobby-Eberly Telescope Dark Energy Experiment \citep[HETDEX;][]{hill2008hobby} Spring Field (right ascension 10h45m00s -- 15h30m00s and declination 45\deg00\arcmin00\arcsec -- 57\deg00\arcmin00\arcsec). LoTSS has a frequency coverage from 120 to 168 MHz, and achieves a typical rms noise level of 70 $\mu$Jy/beam over the DR1 region, with estimated point source completeness of 90 per cent at a flux density of 0.45\,mJy. LOFAR's low frequencies combined with high sensitivity on short baselines gives it high efficiency at detecting extended radio emission. LoTSS DR1 has an angular resolution of 6 arcsec and an astrometric precision of 0.2 arcsec, making it robust for host-galaxy identification. 

In LoTSS DR1 the source detection was performed using the Python Blob Detector and Source Finder \citep[\texttt{PyBDSF},][]{Mohan2015pybdsf}, where a total of 325,694 \texttt{PyBDSF} sources were extracted with a peak detection above 5$\sigma$. \texttt{PyBDSF} fits Gaussians to pixel islands assigning one or multiple Gaussians to each \texttt{PyBDSF} source. The radio catalogues with the \texttt{PyBDSF} properties for both the sources and the Gaussians include positions, angular sizes and orientations, peak and integrated flux density as well as statistical errors.

\texttt{PyBDSF} sources do not always represent true radio sources (i.e. physically-connected sources). Some of the radio components of extended sources may appear as separated and unrelated \texttt{PyBDSF} sources, which need to be associated together into the same source in post-processing. We refer to these as \textit{multi-component} sources in the rest of the paper and they account for 2.8 per cent of LoTSS DR1. In other cases Gaussians may be incorrectly grouped into one \texttt{PyBDSF} source when they are actually distinct physical sources. In this case we refer to them as \textit{blended} sources and they made up only 0.3 per cent of LoTSS DR1. In the vast majority of cases (96.9 per cent in LoTSS DR1), however, \texttt{PyBDSF} correctly associates the radio emission into true physical sources. We refer to these hereafter as \textit{single} sources. These are, in most cases, compact sources composed of only one Gaussian, but can also be extended sources composed by various Gaussians (hence our definition of singles is not the same as the `S' code from the  \texttt{PyBDSF} software used in \citetalias{williams2019lofar}). Even for these correctly associated sources, however, cross-matching with other surveys using statistical means alone can fail due to an incorrect (or missed) counterpart identification, especially if the source is extended and/or asymmetric. This is the case for 1.8 per cent of the sources of LoTSS DR1.

In order to enhance science quality, as part of LoTSS DR1, considerable effort was undertaken to properly associate the radio source components (or dissociate blended sources) and get the correct optical/near-infrared counterparts \citepalias{williams2019lofar}. 
For the majority of LoTSS DR1 sources, \texttt{PyBDSF} correctly associates source components, and outputs an accurate estimate of the position and radio source properties, and therefore such sources were cross-matched statistically using LR. However, complex sources with multiple components or extended emission, and incorrectly blended sources, were sent to visual inspection. This was carried out on a private LOFAR Galaxy Zoo (LGZ) project, hosted on the Zooniverse platform\footnote{\hyperlink{https://www.zooniverse.org}{https://www.zooniverse.org}}, in which each source was inspected by at least 5 collaborators of the LOFAR consortium. The selection of the sources to be analysed in LGZ was done using a decision tree (also referred to as \textit{flowchart}) built using the characteristics of the \texttt{PyBDSF} sources and Gaussians, the neighbouring sources, and the LR of any optical/IR cross-match \citepalias[see][]{williams2019lofar}.  

The decision tree generates 3 main outcomes: the source association and/or identification requires \textit{LGZ}; the source has been correctly catalogued by \texttt{PyBDSF} and the cross-identification (or lack of) can be made by \textit{LR}; and the source is sent to a quick visual sorting (\textit{prefiltering}), where one expert inspects the source and redirects it to one of the other two categories or identifies it as an artefact.
A summary of the number of sources in each of these categories is given in Table~\ref{tab:FC_groups}, where we include in the prefiltering category 223 sources with large optical IDs that were automatically matched to a nearby (large angular size) SDSS or 2MASX galaxy since they were afterwards visually confirmed.
We further exclude 2,591 \texttt{PyBDSF} sources identified by \citetalias{williams2019lofar} as artefacts, except for one source which was automatically marked by the decision tree as an artefact but was instead noted during the LGZ process to be a genuine source. In LoTSS DR1 the artefacts were either removed in an initial stage of the selection process (the majority by being in the proximity of bright sources; 31 per cent) or by visual inspection (mainly during the prefiltering step; 55 per cent). In the next LoTSS releases the improved calibration and imaging pipeline for the radio data \citep[][]{tasse2021lofar} means that we expect a lower proportion of artefacts, most of which will be clearly identifiable and removed at early stages. Furthermore, the properties of any remaining artefacts may be different due to calibration changes. For these reasons we exclude the artefacts when constructing the ML classifier and analysing the results; our final data catalogue therefore contains 323,103 \texttt{PyBDSF} sources. 
The values quoted in Table~\ref{tab:FC_groups} refer to \texttt{PyBDSF} sources and are different to the ones presented on Table 5 of \citetalias{williams2019lofar} which summarises the total number of sources after component association or dissociation.

Using the decision tree, \citetalias{williams2019lofar} initially classified 91.37 per cent of the sources (295,225) as being suitable for LR analysis (see Table~\ref{tab:FC_groups}) and 8.63 per cent (27,878 sources) as requiring visual inspection (either prefiltering or LGZ). These numbers correspond to sources after removal of artefacts. After visual analysis and processing of the final DR1 data, in hindsight, the conclusion is (see Sec.~\ref{sec:classes}) that 95.13 per cent (307,352) could be cross-matched using LR and 4.87 per cent (15,751) required visual inspection. 
For the sources that were sent directly to LGZ (8,195 \texttt{PyBDSF} sources), an examination of the final LGZ decision indicates that 5,051 of them (61.64 per cent) were not correctly associated by \texttt{PyBDSF} and therefore, could not have had their optical identification assigned statistically by LR (or lack of identification in case of no LR match). Similarly, the prefiltering step corresponds to 19,683 \texttt{PyBDSF} sources for which 9,604 \texttt{PyBDSF} sources (48.79 per cent) could not have been processed using LR. In contrast, from the 295,225 \texttt{PyBDSF} sources selected as suitable for cross-matching with LR, 294,129 of them (99.63 per cent) retain the LR cross-match in the final catalogue. In reality, the number of these that are correct will actually be marginally lower since these sources were not subjected to visual examination unless they were part of a multi-component source (usually the core of a radio source) for which one of the source components was sent to visual analysis. This was the method through which the 1,096 sources, sent by the decision tree to LR but which required visual analysis, were discovered. We discuss this in more detail in Sec.~\ref{sec:corrections}.

\begin{table}
 \caption{For each of the main categories (LR, LGZ and prefiltering) classified by the \citetalias{williams2019lofar} decision tree, the table gives the number of sources that were suitable for LR and the number that required visual analysis, as determined using the final outcomes after visual inspection. The final column indicates the percentage of the time that the flowchart decision was correct (i.e. the proportion of sources that were assigned correctly to each of the categories).}
 \centering
 \begin{tabular}{lrrrr}
  \hline
 \citetalias{williams2019lofar}  & Total  & No. suitable  & No. requiring    & Percentage \\
 al. decision & Number & for LR & visual analysis  & Correct\\
  \hline
  LR  & 295,225 & 294,129 & 1,096$^*$  & 99.63 \\
  LGZ & 8,195 & 3,144 & 5,051 &  61.64\\
  Prefiltering & 19,683 & 10,079 & 9,604 & 48.79\\
  \hline
  Total  & 323,103 & 307,352 &  15,751 & 95.57 \\
  \hline
 \end{tabular}
 \smallskip
 \label{tab:FC_groups}
 \parbox{\columnwidth}{$^*$The 1,096 sources selected by the decision tree for LR, but identified as requiring visual analysis, represent a lower limit to the true number as these were only identified when they were part of multi-component sources for which other components were sent to LGZ (see Sec.~\ref{sec:corrections} for further discussion of this).}
\end{table}

\begin{figure}
    \centering
    \includegraphics[width = 1.0
    \columnwidth]{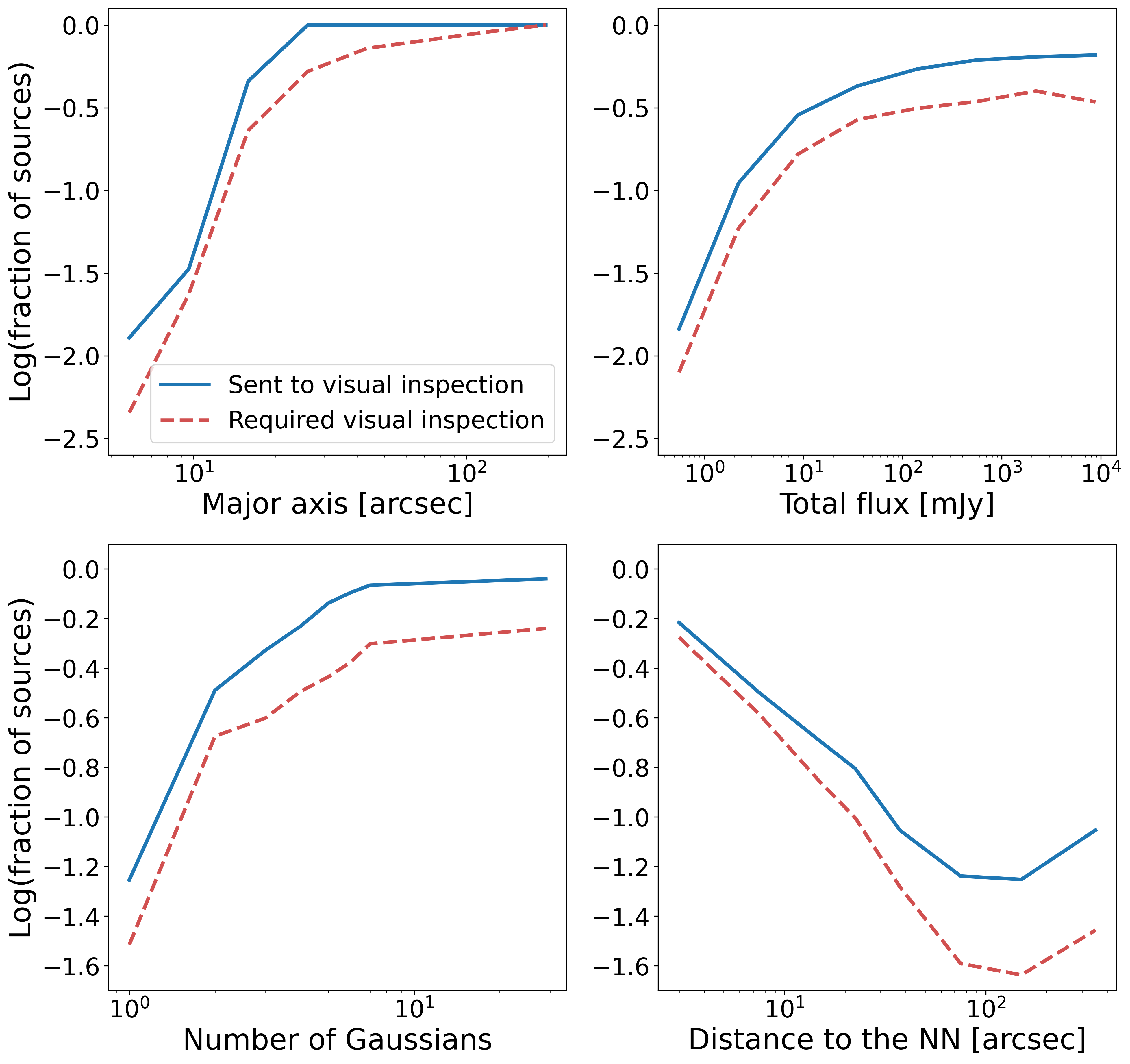}
    \caption{Fraction of \texttt{PyBDSF} sources sent to visual inspection by the \protect\citetalias{williams2019lofar} decision tree (blue lines) and actually required to be inspected (as determined from the final visual inspection outcomes; red dashed lines) 
    as a function of different source parameters: major axis length, total flux density, the total number of Gaussians that compose each \texttt{PyBDSF} source, and distance of each \texttt{PyBDSF} to its nearest neighbour (NN) \texttt{PyBDSF} source.}
    \label{fig:fraction_lgz}
\end{figure}

It is evident from Table~\ref{tab:FC_groups} that overall the \citetalias{williams2019lofar} decision tree has a high accuracy (95.57 per cent). This is mainly because most of the sources are compact and can be cross-matched by LR (where the application of statistical methods results in very high precision). However, the decision tree places about twice as many sources in to the LGZ and prefiltering categories as required, increasing the burden on visual analysis.
Fig.~\ref{fig:fraction_lgz} illustrates the dependence of the decision tree outcomes on some key \texttt{PyBDSF} source properties: the major axis length, the total radio flux density, the number of Gaussians that compose a \texttt{PyBDSF} source, and the distance of each \texttt{PyBDSF} source to its nearest neighbour (NN). In each panel, the blue line shows the fraction of sources that were sent to visual inspection, and the red dashed line shows the fraction of sources that actually needed to be inspected, as determined from the final cross-matched catalogues incorporating the LGZ outcomes. The plots show that the fraction sent for visual analysis increases with increasing source size (note that 15 arcsec was the limit used by \citetalias{williams2019lofar} to distinguish between `small' and `large' sources, with all the large sources being visually inspected, either directly in LGZ or during the prefiltering stage), increasing flux density, increasing number of Gaussian components, and decreasing distances to the NN. These are in line with expectations, as they are all indications that a given source is more likely to be extended and complex. Interestingly, in all cases the red lines are broadly scaled down from the fractions sent to LGZ by about a factor of 2 with no strong parameter dependencies (fluctuations range only from around 1.5 to 2.5 across the parameter space). This indicates that it would not be straightforward to improve the decision tree outcomes simply by adjusting these parameter values. 

\section{Dataset}
\label{sec:machine_learning}

In supervised ML, models are learned from a set of labelled examples drawn from the \textit{dataset}.
The goal is to predict to which \textit{class} a previously unseen example belongs based on the value of its \textit{features}. 
The dataset is a key input for training the ML model and relies upon an adequate and well-profiled number of examples. We create our dataset by evaluating all 323,103 \texttt{PyBDSF} sources from LoTSS DR1 based on their individual characteristics and assigning them to different classes (Sec.~\ref{sec:classes}).
We create different sets of features by using radio source parameters and optical information (Sec.~\ref{sec:parameters}), and we address the class imbalance problem by exploring different ways of balancing the dataset (Sec.~\ref{sec:valid_train_test}). The impact of these last two factors on the classification is investigated further in Sec.~\ref{sec:experiments}. 

\subsection{Classes}
\label{sec:classes}

To create the classes we first evaluate each \texttt{PyBDSF} source (after the results of any deblending or LGZ source association) and assigned them an `association flag' according to different outcomes: the ones that were neither deblended nor associated with other \texttt{PyBDSF} sources (\textit{singles}, flag 1); sources that were deblended (\textit{blends}, flag 2); and \texttt{PyBDSF} sources that were grouped with other \texttt{PyBDSF} sources 
(\textit{multi-components}, flag 4). Note that a small number of sources have a combination of flags since they were first deblended and afterwards one or more of the deblended components was grouped with another \texttt{PyBDSF} source (leading to flag 6).

To create these outcomes, the correspondence between each \texttt{PyBDSF} source and the final radio source association (or lack of association) was assessed using the \texttt{PyBDSF} radio source catalogue from \cite{Shimwell2019lofar} and the final value-added catalogue (source associations and optical IDs) from \citetalias{williams2019lofar}. \texttt{PyBDSF} sources that were grouped with other \texttt{PyBDSF} sources appear as components of a radio source in the corresponding component catalogue, and \texttt{PyBDSF} sources that were deblended appear as two or more radio sources. 

To create a final diagnosis, we also inspected the `single' sources (i.e. the ones with the association flag 1) in order to evaluate whether the LR was a suitable method to identify the host galaxy. This is the case for those sources where the final ID in the value-added catalogue is the same as would have been drawn through LR analysis, or where there was no ID in the final catalogue and the LR analysis also predicted no ID. In contrast, if visual analysis resulted in a change in optical ID (or a change from having no LR ID to having an ID, or vice versa) then these sources are not suitable for cross-matching using the LR method. 
As a result of this evaluation, the sources were assigned into 2 classes (denoted by the flag `accept$\_$lr' throughout this work):

\newcommand\itemLR{\item[\textbf{Class 1:}]}
\newcommand\itemLGZ{\item[\textbf{Class 0:}]}
\begin{enumerate}
    \itemLR \noindent \texttt{PyBDSF} sources that were not associated with other \texttt{PyBDSF} sources, and were not deblended, and for which LR gave the same outcome as was finally accepted in the value-added catalogue (i.e. same host galaxy ID, or correctly gave no ID). These sources would be suitable for LR analysis.
    \itemLGZ \noindent \texttt{PyBDSF} sources that were either associated with other \texttt{PyBDSF} sources in LGZ, or deblended into more than one source, or LR would obtain an incorrect ID. These sources are all unsuitable for analysis by LR alone.
\end{enumerate}

The classes comprise 307,352 sources suitable for LR (class 1) and 15,751 that require visual analysis (class 0); from the latter 9,072 are multiple component \texttt{PyBDSF} sources, 857 are blended \texttt{PyBDSF} sources, and 5,822 are single sources for which a simple application of LR would produce an incorrect ID.
Artefacts (which we  exclude from the analysis) correspond to \texttt{PyBDSF} sources that are not in the final DR1 value-added catalogue.

\subsection{Features}
\label{sec:parameters}

As input features for the ML classifier we used radio source parameters along with properties of the LR matches for both the \texttt{PyBDSF} source being considered and its nearest neighbour (NN). We discuss these below and list them in Table \ref{tab:features}. 

The radio features were built from the \texttt{PyBDSF} catalogue from \cite{Shimwell2019lofar}, where each \texttt{PyBDSF} source has an identifier (Source$\_$Name) with the corresponding radio properties; here we use the major and minor axis sizes and the peak and total flux densities. In addition to these basic radio properties, we used the LR value of the best match and the distance to this match. We computed the LR values for the \texttt{PyBDSF} sources and for each of the Gaussians that comprise a \texttt{PyBDSF} source in the same way as described in \citetalias{williams2019lofar}, with minor modifications that resulted from improvements of the original code \citep{kondapally2021}. 

We also used the Gaussian component catalogue \citep[described in][]{Shimwell2019lofar}, which contains the radio information for all the Gaussians that compose each \texttt{PyBDSF} source. We use the number of Gaussian components comprising a source (indicative of the morphological complexity of the source), and also use the properties (major and minor axis size, fractional source flux density, and LR match properties) of the Gaussian with the highest LR value, or of the brightest Gaussian if the LR of all Gaussian components is below the LoTSS DR1 LR threshold adopted in \citetalias{williams2019lofar}. 

We also used the radio and LR properties of the NN source. In addition, we computed the number of radio sources within 45\arcsec\ (used as an estimate of the local source density, which might be indicative of the presence of multi-component sources) and the flux ratio between the source and its NN. 

Finally, we investigated using the positions of the LoTSS DR1 sources on a cyclic Self-Organising Map \citep[SOM;][]{mostert2021unveiling} as input features. The SOM provides information of the different LoTSS DR1 morphological source `prototypes' on a two dimensional grid. 

In ML, the quality of the features affects the ability of the model to learn. In order to feed useful features that can be more easily interpreted by the algorithm, we made the following transformations to the data: 
\begin{itemize}
    \item We searched the catalogues for missing values (e.g. LR values where there was no potential host within the 15 arcsec search radius) which we assigned extreme values (e.g. a very low arbitrary value of $10^{-135}$ in the case of LR), even thought the tree models adopted in Sec.~\ref{sec:experiments} can in general handle missing data well. 
    \item We used the log value of the number of Gaussians, since complex sources can be made of dozens of Gaussians (up to 51 in LoTSS DR1).
    \item  We encode the values of the SOM morphological prototypes into cyclical features. The prototypes are located on a square grid with (x,y) coordinates. Each radio source is mapped to the prototype of the SOM that it mosts resembles. We transformed the corresponding (x,y) coordinates by using a sine and a cosine transform: this creates 2 new features from each of the original ones, but ensures the cyclical nature of the SOM is retained. We set the values of the prototypes to an arbitrary high value of $10^{20}$  when the source is not available.
    \item We set the value of the LR to a log scale, although this choice has no effect on our results (decision tree models, which we adopt in Sec.~\ref{sec:experiments}, are not sensitive to feature transformations). Using a log scale allows this feature to be used interchangeably with different classifiers (e.g. neural networks).
    \item We create a feature which uses either the LR of the source or the LR of the Gaussian component with the highest LR value if the LR of the source is lower than the LR of one of the Gaussians that make up the source. This is more indicative of a LR match when the source is composed by multiple Gaussians, one of which traces the radio core (and it is the same if the source is only composed by one Gaussian). This can also be indicative of a blended source, especially if the source LR value is below the LR threshold.
    \item We further scaled the LR values by dividing them by the LR threshold value used to process the sources in the HETDEX field (only sources for which the match had a LR value higher than 0.639 got an ID or no ID via this method). This has the advantage of making the model appropriate for future LoTSS fields which might use different optical/near-IR data sets with a correspondingly different LR threshold.
\end{itemize}

\begin{table}
 \caption{List of features used in the analysis. These were selected or calculated using different LoTSS DR1 catalogues$^*$.
 The LR threshold value adopted in LoTSS DR1 ($L_{\text{thr}}$ = 0.639) was used to scale LR value features (these have the suffix tlv). Features in logarithmic scale appear with the log prefix. Sources refer to \texttt{PyBDSF} sources.}
 \label{tab:features}
 \begin{tabular}{ll}
  \hline
  Features & Definition \& Origin\\
  \hline
  \textbf{Baseline (BL)}\\
  Maj & Source major axis [arcsec]$^a$ \\
  Min &  Source minor axis [arcsec]$^a$\\
  Total$\_$Flux & Source integrated flux density [mJy]$^a$\\
  Peak$\_$Flux & Source peak flux density [mJy/bm]$^a$\\
  log$\_$n$\_$gauss & No. Gaussians that compose a Source$^b$\\
  \hline
  \textbf{Likelihood Ratio (LR)}\\
  log$\_$lr$\_$tlv & Log$_{10}$(Source LR value match/$L_{\text{thr}}$)$^c$\\
  lr$\_$dist & Distance to the LR ID match [arcsec]$^c$\\
  \hline
  \textbf{Gaussians (GAUS)}\\
  gauss$\_$maj & Gaussian major axis [arcsec]$^b$\\
  gauss$\_$min & Gaussian minor axis [arcsec] $^b$\\
  gauss$\_$flux$\_$ratio & Gaussian/Source flux ratio$^{a,b}$\\
 log$\_$gauss$\_$lr$\_$tlv &  Log$_{10}$(Gaussian LR/$L_{\text{thr}}$)$^c$\\
  gauss$\_$lr$\_$dist & Distance to the LR ID match [arcsec]$^c$ \\
  log$\_$highest$\_$lr$\_$tlv & Log$_{10}$(Source or Gaussian LR/$L_{\text{thr}}$)$^c$\\
  \hline
  \textbf{Nearest Neighbour (NN)}\\
  NN$\_$45 & No. of sources within 45\arcsec$^a$ \\
  NN$\_$dist & Distance to the NN [arcsec]$^a$ \\
  NN$\_$flux$\_$ratio & NN flux/Source flux density ratio$^a$ \\
  log$\_$NN$\_$lr$\_$tlv & Log$_{10}$(LR value of the NN/$L_{\text{thr}}$)$^c$\\
  NN$\_$lr$\_$dist & Distance to the LR ID match [arcsec]$^c$\\
  \hline
  \textbf{Closest prototype (SOM)}\\
  10x10$\_$closest$\_$prototype$\_$x1  & $\cos$(2$\pi$ Closest$\_$prototype$\_$x/10)$^d$\\
  10x10$\_$closest$\_$prototype$\_$x2  & $\sin$(2$\pi$ Closest$\_$prototype$\_$x/10)$^d$\\
  10x10$\_$closest$\_$prototype$\_$y1  & $\cos$(2$\pi$ Closest$\_$prototype$\_$y/10)$^d$\\
  10x10$\_$closest$\_$prototype$\_$y2  & $\sin$(2$\pi$ Closest$\_$prototype$\_$y/10)$^d$\\
  \hline
 \end{tabular}
  \parbox{\columnwidth}{$^*$ a - \texttt{PyBDSF} radio source catalogue \citep{Shimwell2019lofar};\\
  b - \texttt{PyBDSF} Gaussian component catalogue \citep{Shimwell2019lofar};\\
  c - Gaussian and \texttt{PyBDSF} LR catalogues \citepalias{williams2019lofar};\\
  d - Self-Organising Map for LoTSS DR1 \protect\citep[SOM;][]{mostert2021unveiling}.}
\end{table}

\subsection{Balancing the dataset}
\label{sec:valid_train_test}

The number of objects in the two classes created previously is heavily imbalanced: class 1 has 307,352 sources while class 0 comprises 15,751 objects. The major problem with imbalanced datasets is the tendency of the model to get specialised in the class with more examples (i.e. to overfit to class 1). For that reason, we explore different ways of creating a balanced dataset by under- and over-sampling \citep[cf.][and references therein]{COLLELL2018330}.

We performed under-sampling of the majority class by extracting a random sample of 15,751 objects from class 1 (which is the number of sources available in class 0).
Under-sampling is the standard method adopted throughout the experiments section (Sec.~\ref{sec:experiments}); we use 31,502 sources, comprising the same number of examples in both classes\footnote{Note that when the SOM features are included, this dataset is reduced to 31,320 objects because a small fraction of the LoTSS DR1 sources (on the borders of the mosaics) do not have SOM information.}. In these experiments we used a training set (used to train the model) of 75 per cent of the dataset, and a test set (used to evaluate the model) of 25 per cent. When performing model selection and optimisation (see Sec.~\ref{sec:opt}) we use a 10-fold cross-validation (CV), otherwise we test and train the models on 10 different randomly sampled datasets and use the mean value as the model performance. 

Since both under- and over-sampling have the potential to affect performance, we conducted experiments to determine which method was the best. 
We created a synthetic training dataset with ADASYN \citep[][]{adasyn}, an adaptive sampling technique that is used to generate synthetic examples of the minority class (class 0) by using the original density distribution of the sources in this class. To avoid data leakage, we re-sampled only 75 per cent of the minority class (11841 sources) and tested on a test set comprising the remaining 25 per cent of these sources (which is balanced as well). The number of sources in the training set before and after re-sampling is 303,386:303,386 for class 1 and 11,841:301,738 for class 0, respectively. 
We compare the performance using the model trained using under- and over-sampling in Sec.~\ref{sec:resampling}.

Finally, it should be emphasized that although both under- and over-sampling techniques aim to create a balanced dataset that can generalise well for the two different classes, the distribution of the sources is inherently highly imbalanced and objects that need to be visually inspected are relatively rare in LoTSS (and other deep radio surveys). For that reason, when applying the model trained on a balanced dataset to the real (imbalanced) data, which we do in Sec.~\ref{sec:implications_for_DR1}, other factors require consideration; we discuss these in detail in Sec.~\ref{sec:threshold}.

\section{Experiments}
\label{sec:experiments}
We start by defining in Sec.~\ref{sec:performance} the metrics that will be used to evaluate the performance of the classifier.
In Sec.~\ref{sec:tpot} we create a baseline model for the experiments. This is a less complex yet still effective model that produces acceptable results but has room for improvement. 
The baseline model was selected using the Tree-Based Pipeline Optimization Tool \citep[\texttt{TPOT},][]{Olson2016TPOT}, and consists of a Gradient Boosting Classifier (GBC); see Appendix~\ref{app:ML} for an overview of the ML models and AutoML tools used.
In order to improve model performance, we examine the impact of adding different sets of features in Sec.~\ref{sec:feature_selection}, and optimising the model hyperparameters in Sec.~\ref{sec:opt}.

\subsection{Performance metrics}
\label{sec:performance}

Accuracy is the most common metric to evaluate the performance of a ML classifier. Accuracy can be given as the percentage of the correctly classified inputs relative to the overall classifications: Accuracy $=$(TP+TN)/(TP+FP+TN+FN), where in our case the numbers of false positives (FP), false negatives (FN), true positives (TP) and true negatives (TN) correspond to:

\newcommand\itemtp{\item[\textbf{TP:}]}
\newcommand\itemtn{\item[\textbf{TN:}]}
\newcommand\itemfp{\item[\textbf{FP:}]}
\newcommand\itemfn{\item[\textbf{FN:}]}
\begin{itemize}
    \itemtp sources correctly classified as suitable for LR methods;
    \itemfp sources that should be visually inspected, but which the classifier deems suitable for LR;
    \itemtn sources correctly classified as requiring visual inspection;
    \itemfn sources that could be done by LR techniques but are being sent by the classifier to visual analysis. 
\end{itemize}

When training and testing with a balanced number of examples in each category, accuracy shows the robustness of the classifier. For our binary classification on a balanced dataset, the classifier returns a probability of the source being able to be accepted by LR (class 1), with the probability of being class 0 (requiring LGZ) being 1 minus this probability. 0.5 is the normal threshold value used to discriminate between the two, and it is the value we adopted when evaluating the results in this section and in Sec.~\ref{sec:model_performance_and_interpretation}.
We do, however, investigate other thresholds in order to evaluate the model applied to an imbalanced dataset in Sec.~\ref{sec:implications_for_DR1}. When evaluating the results we are mainly concerned with minimising the number of sources wrongly accepted through LR (FP), while keeping a low number of sources that need to be sent to visual analysis (FN and TN sources).
That is another reason why we further investigate `threshold moving' in order to establish more suitable cut-off probabilities.

We also analyse the values of recall (also known as sensitivity or True Positive Rate; TPR) and precision for our two classes. Precision can be defined as the fraction of sources predicted as being from a certain class that are actually from that class (e.g. TP/(TP+FP)), and recall as the fraction of sources from a certain class that are predicted correctly (e.g. TP/(TP+FN)). The overall balance between precision and recall for the different classes is given by the F1-score (2$\times$(Precision$\times$Recall)/(Precision+Recall)). In Sec.~\ref{sec:model_performance_and_interpretation} we also use the False Positive Rate (FPR) to illustrate the performance of the classifier. The FPR corresponds to the fraction of sources from class 0 that are incorrectly classified (FP/(FP+TN)).

In our analysis we further define the `LGZ scale-up factor' which corresponds to the total number of sources that we would have to visually inspect scaled by the ones we should really inspect ((FN+TN)/(TN+FP)). In other words, it represents the multiplicative factor of additional galaxies we would have to send to LGZ besides the ones that should be sent.
We compare it with the False discovery rate (FDR = FP/(FP+TP)), which corresponds to the fraction of sources deemed to be suitable for cross-match by LR that are classified incorrectly.

\subsection{Baseline}
\label{sec:tpot}
In order to create a baseline model, we used \texttt{TPOT} and a set of baseline features (BL) that contain only basic radio source information: \texttt{PyBDSF} peak and total fluxes, major and minor axis sizes, and the logarithm of the number of Gaussians that compose each \texttt{PyBDSF} source.

We ran \texttt{TPOT} using a set of conservative parameters: 3 generations (number of iterations of the optimisation process; see Appendix~\ref{app:ML} for more details), a population size of 20 (number of candidate solutions \texttt{TPOT} retains in each generation), and a 10-fold CV (number of data splits where each pipeline is trained and evaluated). This allows \texttt{TPOT} to search for 600 different models in each run. The choice of the values for these parameters is subjective, and higher values would enable the search for more model combinations. However, running \texttt{TPOT} for a larger number of generations and population size would drive \texttt{TPOT} towards more complex ML pipelines with stacked models that could cause the model to overfit; this is a current challenge of the method \citep[see][for a discussion]{Olson2016TPOT}. Therefore, we define low values for the \texttt{TPOT} parameters, and use it to get recommended pipelines. In that way, we select a simple model that provides interpretability for our experiments and we perform model optimisation at a later stage.

We performed different \texttt{TPOT} runs and we found a consistent selection of tree-based models as the favoured choice: using different balanced random samples of the full dataset, \texttt{TPOT} would select a GBC or occasionally an XGBoost (XGBoost is an optimised version of a GBC which can include regularisation and allows further optimisation due to the amount of parameters that can be tuned); when using subsets of the data (half-size dataset) a Random Forest or an Extra Trees classifier was favoured. For all the models we achieved an internal CV accuracy of around 89 per cent and a test accuracy within $\pm$0.5 per cent of the CV value. The GBC achieved higher performance on the CV tests but the Random Forest models showed a higher generalisation ability when training with only 50 per cent of the dataset. This suggests that the smaller dataset does not contain enough examples for \texttt{TPOT} to detect strong patterns among the features and therefore it fits a model that performs well with higher variance data.
This also indicates that the classification could benefit from adding more relevant features and could be improved using a GBC model (with optimised hyperparameters, such as a bigger ensemble size and/or a different learning rate). For our baseline model, we therefore select a GBC with 100 estimators and a learning rate of 0.01, which are also the hyperparameters suggested by \texttt{TPOT}. The complete specifications of the baseline model can be seen in Table~\ref{tab:hyperparameters}.

\begin{table}
 \caption{Accuracy on the test sets of the GBC model before and after optimisation of the hyperparameters, and using cumulative sets of features: baseline features (BL), \texttt{PyBDSF} source LR features (LR), \texttt{PyBDSF} Gaussian features (GAUS), nearest neighbour (NN) and SOM features (SOM) as described in Table~\ref{tab:features}. In each case the GBC was run on 10 different down-samplings with random sampling of the dataset into training and test sets, and the mean of these 10 is quoted. The standard deviation between the 10 datasets is typically around 0.2 per cent.}
 \begin{tabular}{lcc}
  Set of features & \multicolumn{2}{c}{Accuracy achieved} \\
  & Baseline hyperparameters & Optimised GBC \\
  \hline
  (0) BL & 88.7\% & 88.7\% \\ 
  (1) BL \& LR  & 90.2\% & 90.2\% \\
  (2) 1 \& GAUS & 90.3\% &  90.3\% \\ 
  (3) 2 \& NN &  94.4\% &  94.6\% \\ 
  (4) 3 \& SOM & 94.7\% &  94.8\% \\ 
  \hline
 \end{tabular}
 \label{tab:experiments}
\end{table}

\subsection{Feature selection}
\label{sec:feature_selection}
We started with the baseline model and investigated the impact of adding different sets of features (as described in Table~\ref{tab:features}) to the classifier; their impact on classification is illustrated in Table~\ref{tab:experiments}. These comprise four sets of features in addition to the (0) baseline features: (1) LR information of the \texttt{PyBDSF} source; (2) properties of the Gaussian component with the highest LR value (or the brightest Gaussian if none have a LR match); (3) the nearest \texttt{PyBDSF} neighbour information, and (4) the positions of the \texttt{PyBDSF} sources on the SOM. 

\begin{figure*}
    \centering
    \includegraphics
    [width = 1 \textwidth]
    {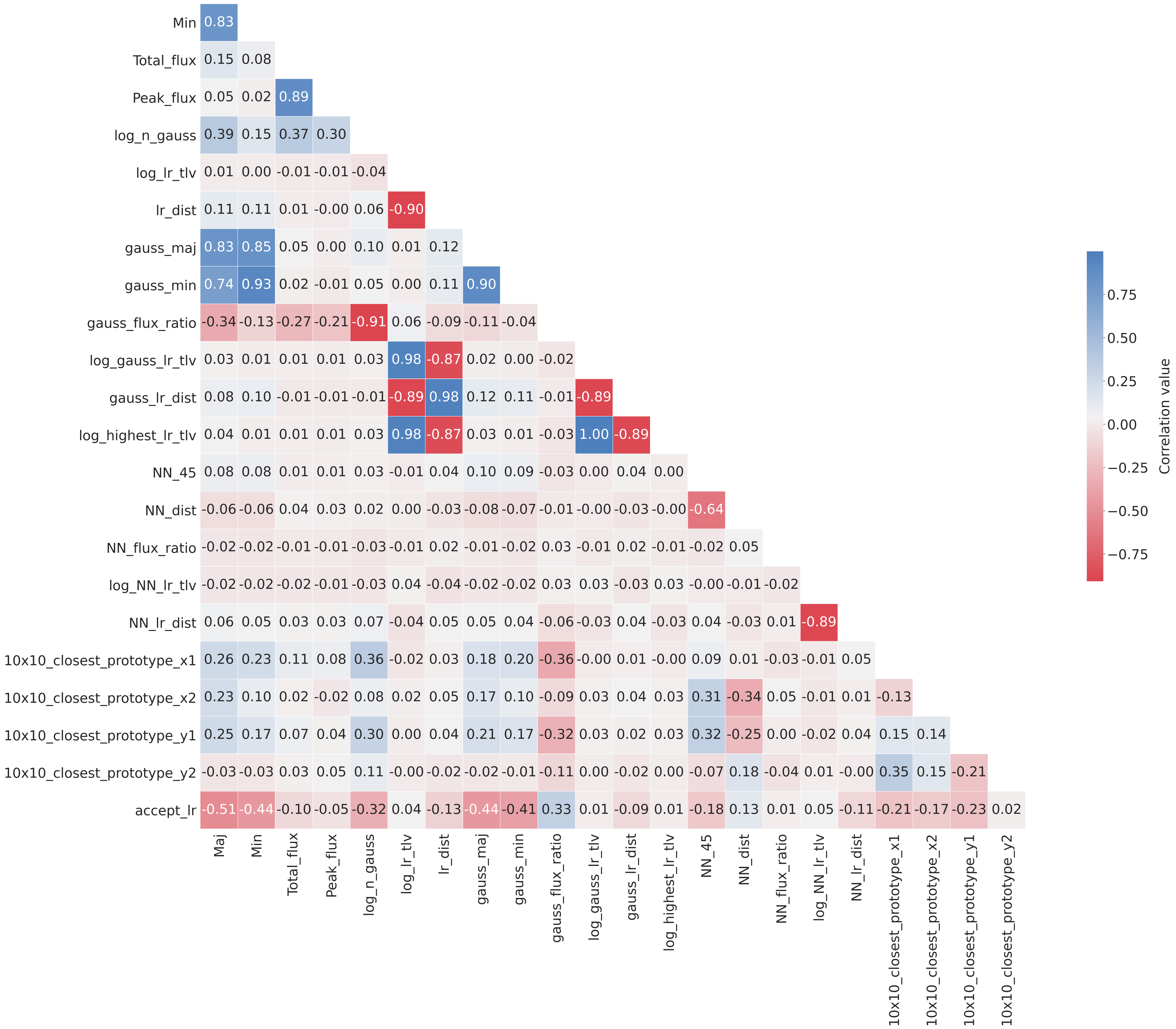}
    \caption{Correlation matrix using Pearson correlation. This shows the correlation coefficients between each of the different input features considered for the modelling (blue for positive linear correlation, red for negative linear correlation, scaling from 1 to -1). The bottom row provides the correlation of each parameter with the final `accept$\_$lr' outcome, indicating the strength of any linear relation between the features and the target class.} 
    \label{fig:feature_correlation}
\end{figure*}

\noindent {\textbf{Source LR features:}}
the addition of the LR features (LR value and LR distance) of the \texttt{PyBDSF} source increases the performance accuracy of the baseline model by about 1.5 per cent (from 88.7 per cent to 90.2 per cent; see Table~\ref{tab:experiments}). This improvement is expected, as the presence or absence of a potential host galaxy at the expected position is a strong indicator of whether the source has been correctly associated. 

\noindent {\bf Gaussian (GAUS) features}: the addition of the Gaussian features has a small impact on the model with only minor improvements for the classification. When adding these features to the Baseline features and the LR features, the improvement is 0.1 per cent.
Fig.~\ref{fig:feature_correlation} shows the correlation between different input features (and with the resulting classification). It is evident from this plot that the flux ratio relative to the source and the size of the Gaussian (gaus$\_$min and gaus$\_$max) do show, respectively, a strong positive and strong negative correlation with the `accept$\_$lr' output, and thus contain useful information. However, the sizes of the Gaussians show a very strong correlation with the sizes of the sources, the flux ratio between the Gaussian and the source is highly correlated (inversely) with the number of Gaussians that composed each \texttt{PyBDSF} source, and the Gaussian LR features (log$\_$gauss$\_$lr$\_$tlv and gauss$\_$lr$\_$dist) are also highly correlated with the source LR parameters (not least because most sources are composed by single Gaussian components). Thus, the inclusion of the Gaussian features does not introduce much new information. Nevertheless, we include these features in our final model as they are easily available and offer marginal improvement.

\noindent {\bf Nearest Neighbour (NN) features:}
adding the NN information has the greatest impact on the model performance, improving the classification by more than 4 per cent. Even though there is not a strong linear correlation with the `accept$\_$lr' output in Fig.~\ref{fig:feature_correlation}, the NN$\_$dist, NN$\_$lr$\_$dist and log$\_$NN$\_$lr$\_$tlv, and the NN$\_$45 parameters provide valuable additional information for the classification, as does the flux ratio of the NN source relative to the source under consideration. 

\begin{table*}
 \caption{Gradient Boosting Classifier (GBC) model hyperparameters: baseline, tuning values, and finally-adopted optimised hyperparameters obtained by \texttt{TPOT} optimisation. The learning rate controls how quickly the loss is corrected at each iteration; no. of estimators corresponds to the number of sequential trees create by the model; max depth represents the maximum tree extension; subsample is the proportion of data used in each tree; min samples split corresponds to the minimum number of examples necessary to split a tree into different branches; min samples leaf is the minimum number of examples required in a terminal leaf; and max features is the maximum number of features to take into consideration while searching for the optimal split.}
 \begin{tabular}{c c c c c}
 Hyperparameters & Baseline GBC & Search values  & Optimised GBC \\
   \hline
   learning rate & 0.01 & 0.001, 0.01, 0.05, 0.1, 0.5, 1 & 0.01 \\
   no. of estimators & 100 & 100, 250, 500, 1000 & 500 \\
   max depth & 10 & range (1, 11, steps = 1) & 8 \\
   subsample & 0.75 & range (0.05, 1.01, steps = 0.05) & 0.15\\
   min samples split & 6 & range (2, 21, steps = 1) & 12\\
   min samples leaf & 10 & range (1, 21,  steps = 1) & 5\\
   max features & 0.35 & range (0.05, 1.01, steps = 0.05) & 0.6\\
 \hline
 \end{tabular}
 \label{tab:hyperparameters}
\end{table*}

\noindent {\bf Self-Organising Map (SOM)} features: experiments using solely the baseline and the SOM features improves the classifier by about 2.5 per cent compared to the baseline only.
Impressively, if using only the SOM as input features (not shown in Table~\ref{tab:experiments}), the model achieves a classification of almost 80 per cent, which demonstrates the power of the morphological representation for the classification. However, it also demonstrates that some essential information contained within the baseline features is not retrievable from the SOM alone.

The addition of the SOM features on top of all of the other different experiments improved the model accuracy by 0.3 per cent, to 94.7 per cent on the baseline model and 0.2 per cent on the final model. 
This indicates that the information encoded in the SOM, through a visual representation of the source (compact vs extended emission, single vs blended vs multiple radio component source, etc) does provide some additional information over the other features. However, this is limited, due to the correlations between the SOM and other features as seen in Fig.~\ref{fig:feature_correlation}.  
Due to the relatively small improvement, and because the SOM features come from an external source, we have decided to exclude the SOM from our final model. 

\noindent {\bf Deconvolved features:} we also investigated using the deconvolved (DC) major and minor axis instead of the measured values, and we found the same results. We ran the model using the DC and non-DC major and minor axis for both the \texttt{PyBDSF} sources and the Gaussians and the differences were negligible. Baseline experiments replacing the measured sizes by the deconvolved sizes of the sources pointed to a small improvement on the classifier, but well within the range of the variance of the model. In our final model we opted to use the non-deconvolved sizes as these are potentially more robust against inaccurately measured beam sizes; however, this choice is arbitrary and is not expected to have a significant effect on the classifier for LoTSS DR1. 

\subsection{Model optimisation}
\label{sec:opt}

\subsubsection{Selection of model and model hyperparameters}

After feature selection we performed further experiments using \texttt{TPOT} to optimise the model hyperparameters using a single dataset. The hyperparameters are used to adjust the learning process (e.g. learning rate) and the model specifications (e.g. number of estimators, i.e. trees, on a tree-based model). We ran \texttt{TPOT} for 3 generations with a population size of 5, and a cross-validation (CV) of 10 and the sets of features from Table~\ref{tab:features} excluding the SOM. The range of values we defined for  \texttt{TPOT} to perform the search, and the optimised set of model hyperparameters for the GBC model finally selected, can be seen in Table~\ref{tab:hyperparameters}.

Since there is some discussion in literature about boosting methods overfitting under certain circumstances (see Appendix~\ref{app:ML} for references) we give special attention to check that the model we use does not overfit. Therefore, and for verification purposes, we tested different possible combinations of hyperparameters. Increasing the learning rate and increasing the number of estimators both make the model increase its accuracy; for example, for 1,000 estimators the accuracy is able to reach values higher than 99 per cent on the training set and 94.8 per cent on the test set. However, a training set performance close to 100 per cent is a strong indication that the model is overfitting, especially with the significant difference in performance between the training and test sets (although the high accuracy on the test set shows the model is still able to generalise). \texttt{TPOT} favours the use of 500 estimators, which offers good results and minimises the risk of over-fitting. Our optimised GBC model achieves an internal \texttt{TPOT} CV score of 94.6 per cent and an average accuracy of 94.6 per cent on the test and 95.9 per cent on the training set.\footnote{Note that this accuracy can not be fairly compared against the accuracy of the decision tree of \citetalias{williams2019lofar} quoted in Table~\ref{tab:FC_groups}, since the latter is for a very unbalanced dataset and is optimised for performance on the majority population of class 1 sources. We compare the ML performance against that of the \citetalias{williams2019lofar} decision tree in Sec.~{\ref{sec:flowchart_comparison}.}} These are also the values obtained for the model trained and optimised using a single dataset which we further use to present the results in the next section. This is within 0.2 per cent of the performance with 1,000 estimators, but by using a smaller number of estimators we reduce the complexity of the model as well as training time, and can have higher confidence that the model is not overfitting. 

We also investigated an XGBoost model, as this was also favoured by \texttt{TPOT}. The best XGBoost model achieves an internal \texttt{TPOT} CV score of 94.6 per cent and an average accuracy of 94.7 per cent on the test and 96.6 per cent on the training set. This is a marginally superior performance on the test set to the GBC model, but within the scatter of different dataset selections, and also has a higher difference between test and training set performance. Given this, we opt to retain the less complex GBC model for our final analysis.

Overall, as can be seen from Table~\ref{tab:experiments} and Table~\ref{tab:hyperparameters}, the hyperparameters and performance for the optimised GBC model are not dissimilar from those of the baseline model.

\subsubsection{Training with re-sampling}
\label{sec:resampling}

To test whether under- or over-sampling is a better approach, we applied the optimised classifier on the re-sampled data (see Sec.~\ref{sec:valid_train_test}). Not surprisingly, we found that training the model with more examples of class 0 (even if they are synthetic), results in a higher precision for this class. Additionally, when compared to training without resampling, it results in a more proportional model performance across the two classes. 
This model reduces the number of sources that need to be visually inspected (the value of recall for class 1 increases) but this comes at the cost of accepting more sources for LR than should be (precision on class 0 decreases). This increase in the number of false positives is not in alignment with our science goals, as these sources will all remain incorrectly classified in the final analysis. 
The overall performance for the re-sampled datasets decreases by 0.7 per cent in accuracy on the test set, compared to the down-sampling method, while the accuracy for the training set increases by 1.16 per cent. This difference is particularly evident for sources in class 1, for which the model got too specialised: it achieves 98.41 per cent precision on the training set, which does not allow it to generalise well on the test set for this class. This is the most probable reason why the model accepts too many false positive sources as suitable for LR analysis. We conclude that training with re-sampling leads to overfitting the classifier, and hence we opt for training the final classifier with down-sampling instead. 

\section{Model performance and interpretation}
\label{sec:model_performance_and_interpretation}

\subsection{Final Model Performance}

\begin{table}
 \caption{Performance on the test and training sets: the results give the overall accuracy, and the F1-score, precision and recall for each class (where 1 $=$ suitable for LR; 0 $=$ requires LGZ), for a decision tree threshold of 0.50  or 50 per cent. The results quoted are for a single down-sampled balanced dataset.}
 
 \begin{center}
 \begin{tabular}{l c c}
   & Test set & Training set \\
  \hline
  Accuracy     & 0.946 & 0.959\\
  F1-score 1   & 0.945 & 0.958\\
  F1-score 0   & 0.947 & 0.960\\
  Precision 1  & 0.963 & 0.975\\
  Precision 0  & 0.930 & 0.944\\
  Recall 1     & 0.928 & 0.942\\
  Recall 0     & 0.964 & 0.976\\
  \hline
 \end{tabular}
 \label{tab:statistics}
 \end{center}
\end{table}

The model that we adopt in the rest of the paper is the GBC model with the optimised hyperparameters described in Table \ref{tab:hyperparameters} and the 18 features (which exclude the SOM features) from Table~\ref{tab:features}, trained and tested on a balanced dataset created with down-sampling. In Table~\ref{tab:statistics} we present the suite of metrics defined in Sec.~\ref{sec:performance} to assess the performance a binary classifier, in order to illustrate the overall performance of the model, as well as the performance on the different classes. The results presented here are run on an independent testing set and adopt a standard cut-off probability of 50 per cent between the two classes.

\begin{figure}
    \centering
    \includegraphics[width = 0.9 \columnwidth]{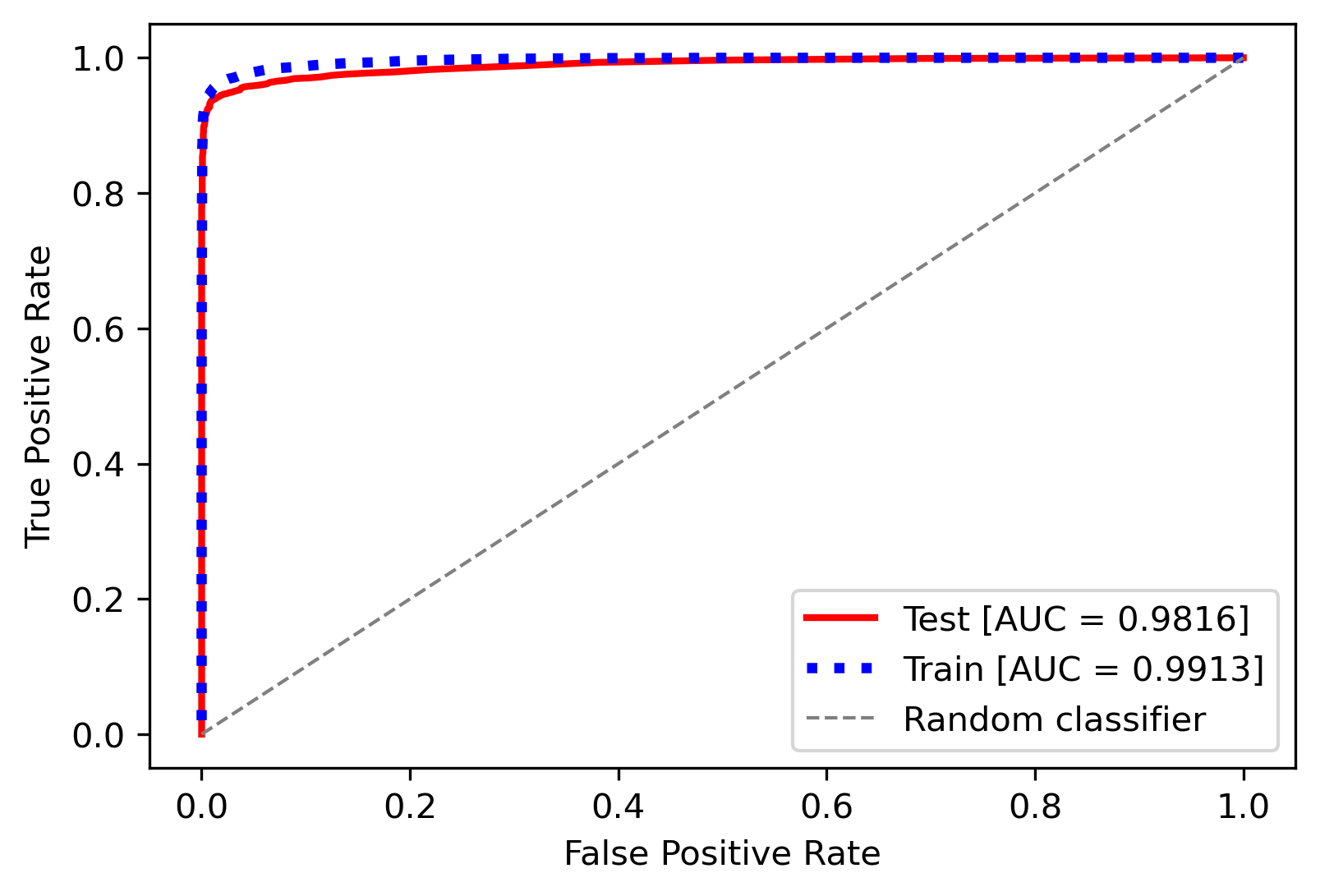}
    \caption{Receiver Operating Characteristic (ROC) curve of the optimised model for a train and test balanced dataset, showing that this has an Area Under the Curve (AUC) close to unity, where 1 would be the value for a perfect classifier classifier. The True Positive Rate (TP/(TP + FN)) is the rate at which a source suitable to cross match with LR is correctly identified as such out of all the ones that can be done using this method, while the False Positive Rate (FP/(TN + FP)) is the proportion of sources that are incorrectly predicted to be suitable to LR out of all the ones that require visual inspection.}
    \label{fig:roc}
    \end{figure}

Our best model achieves an overall accuracy of 94.6 per cent on the testing set, and just 1.3 per cent higher on the training set. The model can be seen to favour precision for class 1 (sources that can be cross-matched using LR) and recall for class 0 (sources that require visual inspection). These are the values we intend to optimise: while we want to avoid a high number of visual inspections it is more important to reduce the number of sources accepted as class 1 when they do not belong to that class. From the total number of sources accepted as being suitable for LR analysis, 96.3 per cent are actually from that class; similarly, 96.4 per cent of the sources that need to be visually inspected are sent to visual inspection. While this means that there is already a low percentage of sources wrongly predicted to be class 1, in practice the number that will end up being mis-classified is even smaller as some of these sources will be corrected during the LGZ process (see corrections applied in Sec.~\ref{sec:corrections}). The model yields slightly lower values of precision for class 0 and recall for class 1, meaning that the model sends more sources to visual inspection than needed. Overall, the classification predictions send around 7 per cent more sources (in a balanced dataset) to visual inspection than needed to be inspected; this percentage will be significantly higher when applying the model to a highly-imbalanced dataset with many more sources in class 1. 

For illustration, we show in Fig.~\ref{fig:roc} the Receiver Operating Characteristic (ROC) curve of the model. This shows the True Positive Rate (TPR) against the False Positive Rate (FPR) plotted for different thresholds. The plot illustrates the performance of the model on detecting a source that can be processed by LR (i.e. a positive test) as we achieve values close to a TPR of 1 and FPR of 0; and an AUC (Area Under the Curve) for the test set of 0.98 (where an AUC of unity would correspond to the perfect classifier). Instead of using the default 0.50 threshold for balanced datasets, we can further explore a more suitable cut-off threshold closer to the top left corner of the curve, which is particularly important when dealing with imbalance datasets. We therefore explore the effect of varying the cut-off threshold in Sec.~\ref{sec:threshold} in order to optimise the trade-off between the number of sources wrongly accepted as suitable for LR and the number of sources sent to visual inspection.

\subsection{Feature Importance in the model}

To interpret the importance of the different features for the classification we use SHAP \citep[SHapley Additive exPlanations; ][]{Lundberg2017}, through the use of a Python package explicitly applied to tree-based ML models \citep{lundberg2020local2global}. 
The method measures the impact of different features on the model classification by averaging the contribution of a particular feature compared to when that feature is absent for the prediction.

\begin{figure*}
\centering
\begin{minipage}{.5\textwidth}
  \centering
  \includegraphics[width=0.9\linewidth]{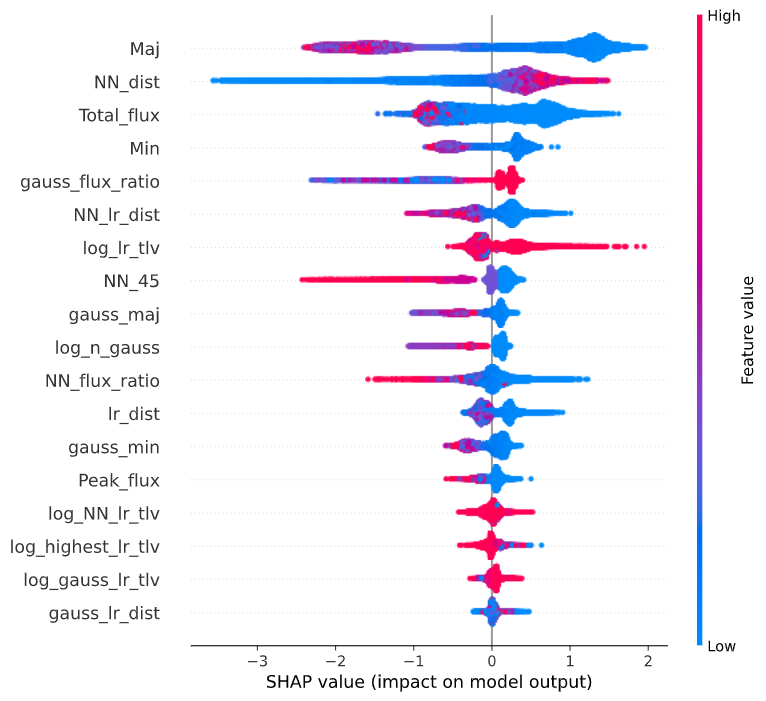}
\end{minipage}%
\begin{minipage}{.5\textwidth}
  \centering
  \includegraphics[width=0.9\linewidth]{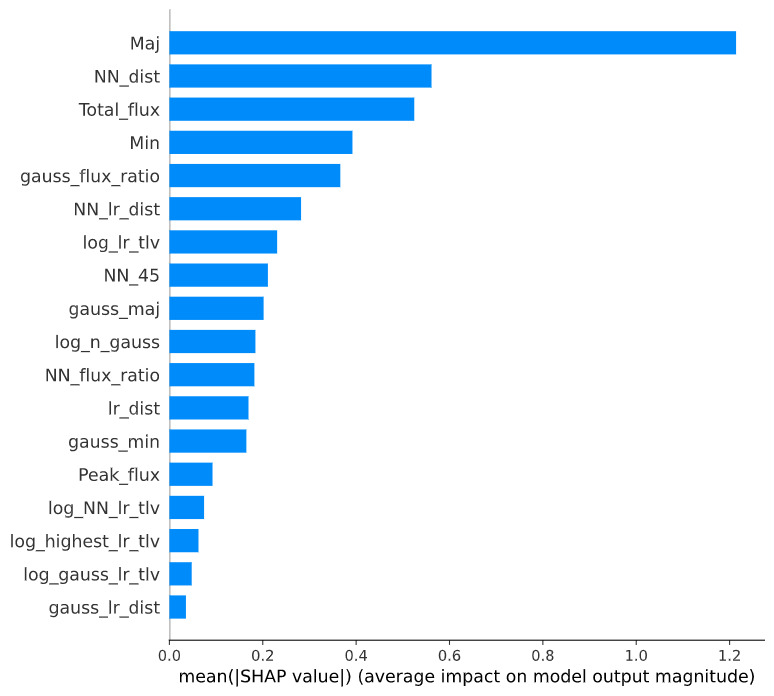}
\end{minipage}
\caption{{\it Left}. SHAP values for each feature and for each source within the training set. The colour coding indicates the value of the feature for that source compared to the range of values for that feature across all sources, as indicated by the colour bar, and the thickness of the plot indicates the density of sources at a given SHAP value. 
Larger absolute SHAP values indicate higher impact in the prediction. {\it Right:} SHAP feature importance computed as the mean of the absolute SHAP values. These are ordered such that the features with the highest predictive power are at the top.}
  \label{fig:feature_importance}
\end{figure*}

The SHAP values are computed individually for each source in the training set, and the left panel of Fig.~\ref{fig:feature_importance} shows how the values of each feature contribute to the classification. SHAP values are given in units of log of odds, with 
positive SHAP values implying that the value of the feature favours class 1 sources and negative SHAP values implying that the feature value favours class 0 sources. The colour-coding on the plot indicates the value of the input feature compared to the range of values of that feature for all sources. Thus, for example, higher values of the major axis are associated with sources that have highly negative SHAP values (class 0), while lower major axis values favour class 1. 

The right-hand panel of Fig.~\ref{fig:feature_importance} shows 
the global contribution of the different features to the model predictions, in descending order. These correspond to the mean of the absolute SHAP values per feature across all the data on the training set. The features at the top of the plot are those with the highest predictive power: these are the major axis of the source followed by the distance to the source's NN. The features towards the bottom of the plot provide the least predictive power of those considered in the model. 
 
\begin{figure*}
    \centering
    \includegraphics[width = 0.85 \textwidth]{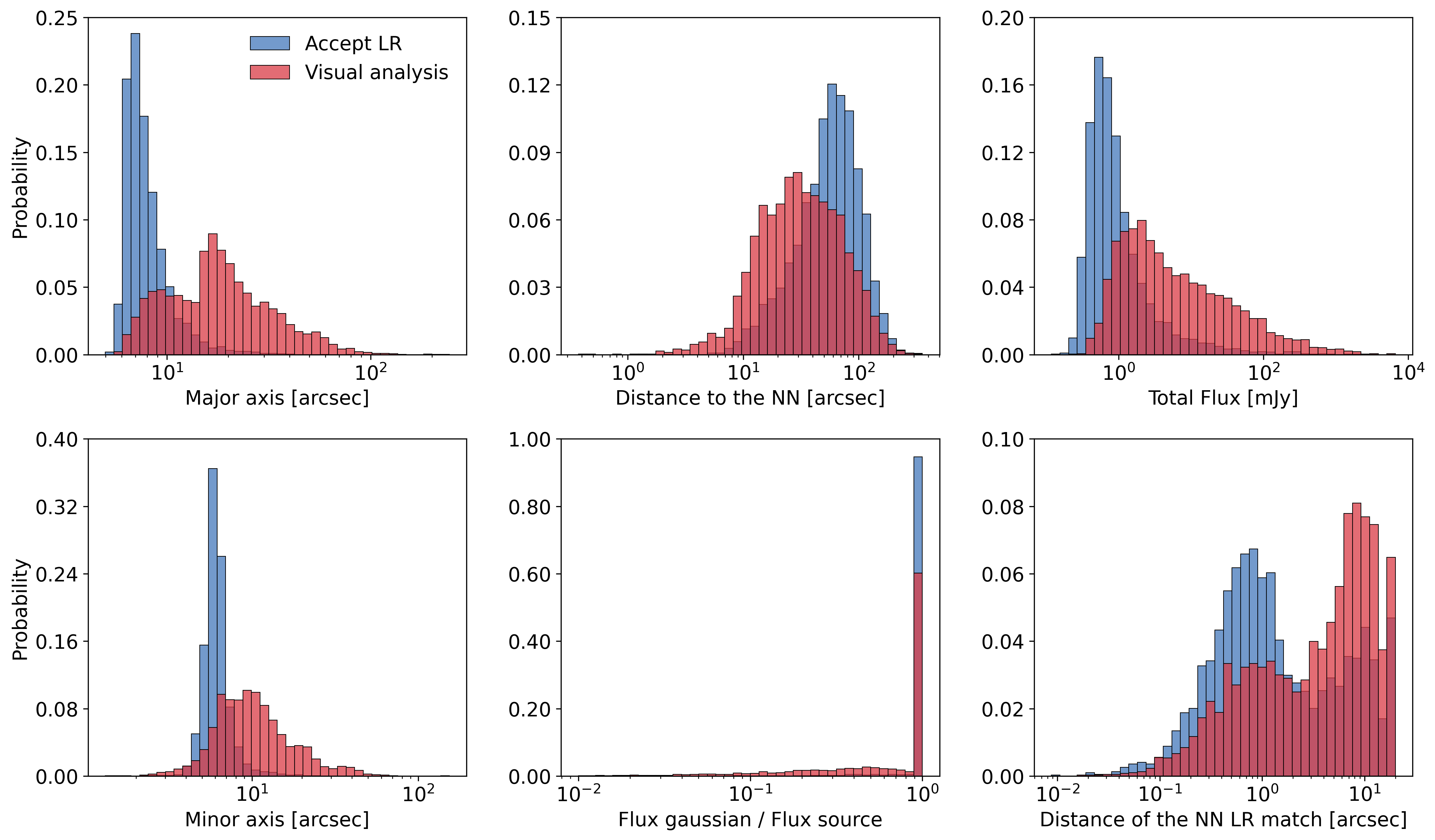}
    \caption{Probability distributions of the most distinctive features, as identified by the SHAP analysis. In each case, blue corresponds to sources that are suitable for statistical match by LR (class 1) and red represents sources that require visual analysis (class 0). For all of these features, a systematic offset in feature values between class 1 and class 0 sources is apparent, in the direction that would be expected from the radio source properties (see text for details).}
    \label{fig:hist}
\end{figure*}

Fig.~\ref{fig:hist} shows the distribution of feature values for the six features picked out to have the highest predictive power. Specifically, it shows histograms of the distributions of feature values for the two classes of objects (class 0, class 1), each normalised to the total number of sources of that class. In each case a distinction between the two classes is apparent, and is in the direction which would be expected. Smaller sources (both major axis in the upper left and minor axis in the lower left) have a higher probability of having a correct cross-match by statistical means, as opposed to more extended sources, which are more likely to be resolved and possibly complex. Brighter sources (upper right) are also more likely to require visual analysis, due to the predominance of more extended AGN at higher flux densities compared to more compact AGNs and SFGs at fainter flux densities \citepalias[see discussion in][]{williams2019lofar}. Sources for which the Gaussian component contains only a fraction of the total flux density and hence other Gaussian components must also be present, indicating an extended source, are also more likely to need visual analysis (lower middle panel), as compared to compact sources with all of their flux in a single Gaussian. Finally, those sources with a close near neighbour (upper middle panel), especially when that near neighbour does not have a close LR match (lower right panel) are also indicative of multi-component sources which require visual analysis.

\section{Application to full LoTSS datasets}
\label{sec:implications_for_DR1}

In this section we apply our model to the full LoTSS DR1 dataset, and also make a preliminary evaluation of its performance on a subset of LoTSS DR2. 
When applying the trained ML model to the full LoTSS DR1 dataset there are two main points that require consideration. First, unlike the dataset used to train and test the model, LoTSS DR1 is highly imbalanced. In Sec.~\ref{sec:threshold} we investigate varying the cut-off probability to select a value that is more suitable for this class distribution problem rather than using the default 0.5 threshold. We also define the parameters by which we will assess the performance of the model in order to select the appropriate threshold. Second, it should be noted that some sources wrongly classified by the algorithm as being suitable for LR (false positives) may be recovered (corrected) if additional components of the same (multi-component) source are sent to LGZ. This may particularly be the case for the cores of extended radio sources: the core itself is compact and aligns with the optical host galaxy so may have a higher LR match, pushing towards a class 1 prediction, but the surrounding extended lobes are far more likely to be predicted to need LGZ. We examine and correct for this issue in Sec.~\ref{sec:corrections}.

To investigate the overall performance of the classifier in different regions of parameter space, we compare our results with those of the \citetalias{williams2019lofar} decision tree in Sec.~\ref{sec:flowchart_comparison} and investigate the success of the classifications for different source properties (as defined from the SOM) in Sec.~\ref{sec:perform_subsets}. Finally, we conclude the evaluation of the model on LoTSS DR1 by examining the nature of those sources that deliver false positive outcomes (i.e.\ are sent to LR but should require LGZ) in Sec~\ref{sec:FP}.
In Sec~\ref{sec:application_to_DR2}, we further apply our model directly to LoTSS DR2 as a first step to evaluate how the model performs in a completely unseen dataset.

\subsection{Threshold moving for an imbalanced dataset}
\label{sec:threshold}

\begin{figure}
\centering
  \includegraphics[width=1\linewidth]{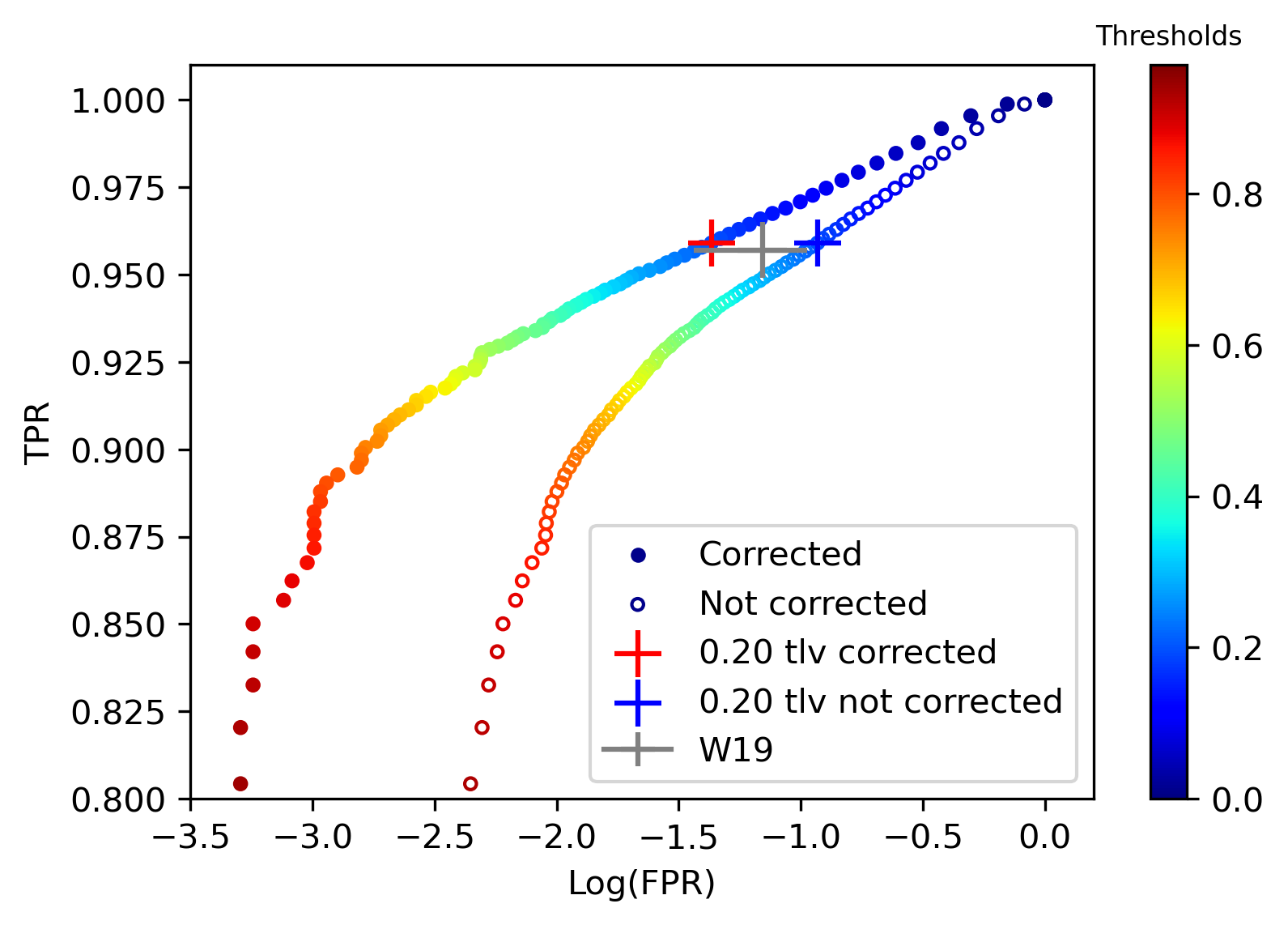}
  \caption{Zoom in of the ROC for the full LoTSS-DR1 dataset, showing the different threshold levels. Note that to better visualise the results, the x-axis is on a log scale, and only the upper values of the y-axis are shown (cf. Fig.~\ref{fig:roc}). The open (lower) symbols represent the raw results from the model fitting, and the filled (upper) symbols demonstrate the improvement which results from the corrections for recovered false positives (see Sec.~\ref{sec:corrections}). The threshold value adopted is indicated by the red and blue crosses which corresponds to a False Positive Rate (FPR) of 11.7 per cent for not corrected and 4.3 per cent for the corrected values, and a True Positive Rate (TPR) of 95.8 per cent. The grey point indicates the results of the \citetalias{williams2019lofar} decision tree using the raw values from Table~\ref{tab:FC_groups}, with the horizontal error bar representing the potential spread from uncorrected to corrected values if the false positive recovery rate for \citetalias{williams2019lofar} would be the same as for the classifier.}
  \label{fig:roc_dr1}
\end{figure}

The distribution of the two classes in the LoTSS-DR1 dataset is severely skewed towards class 1, and the default 0.5 threshold value does not represent an optimum cut-off probability between the two classes. The model prediction threshold reflects the proportion of examples in the two classes that were used to train the classifier; as a result, when the model is applied to the entire, imbalanced LoTSS-DR1 dataset, the majority of sources are classified as belonging to class 1, which is the most frequent class. Therefore, we tune the decision threshold, often known as `threshold moving', which is a common approach used to optimise the predictions for imbalanced datasets \citep[e.g.][]{COLLELL2018330}.
The effect of changing the threshold is demonstrated on the ROC curve in Fig.~\ref{fig:roc_dr1}.

\begin{figure}
\centering
  \includegraphics[width=1.0\linewidth]{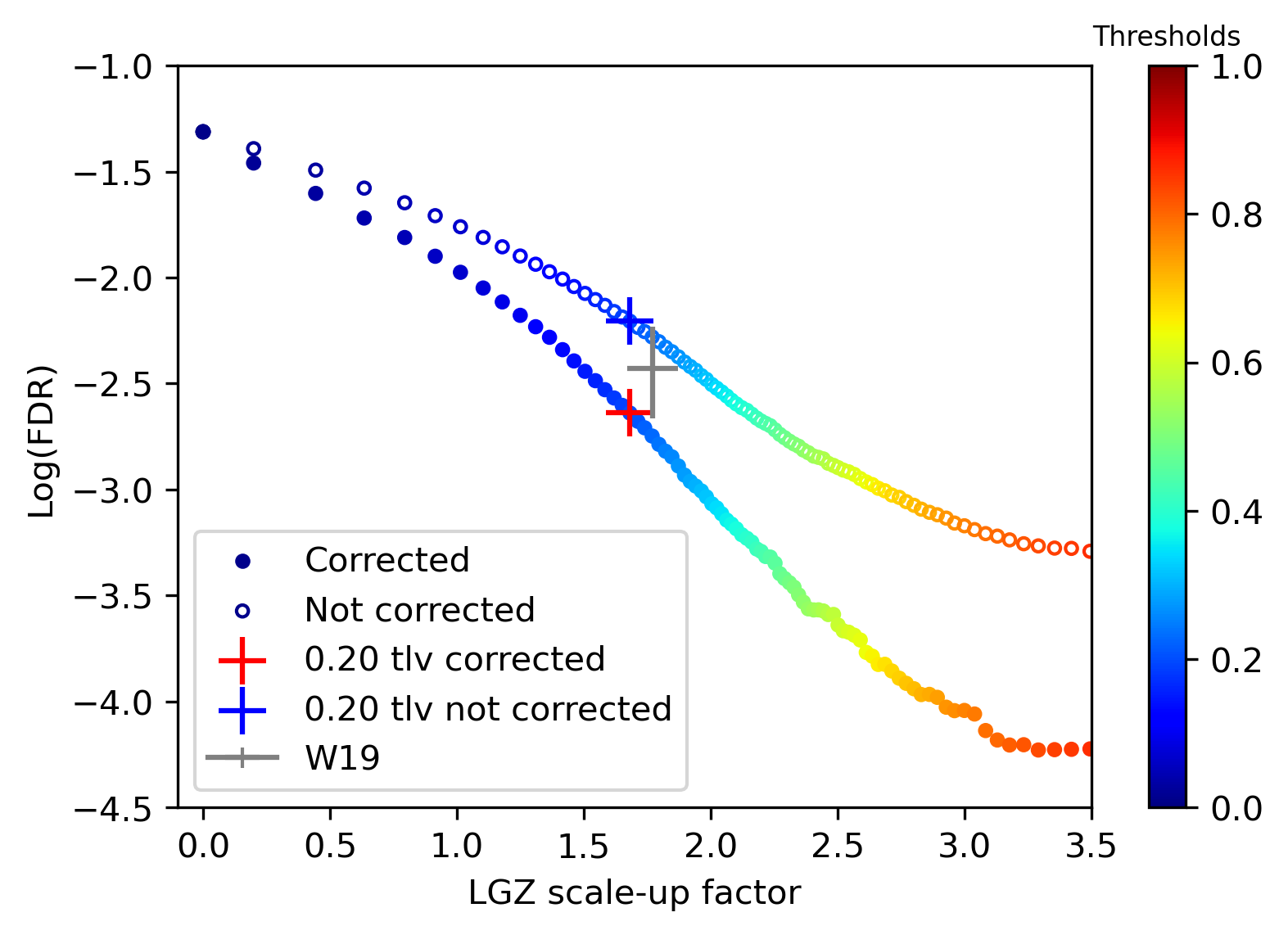}
  \caption{A comparison of the performance metrics adopted for analysis of our model for different values of the cut-off threshold between the two classes. The y-axis is the False Discovery Rate (FDR), which measures the fraction of sources accepted for LR which were incorrectly selected. The x-axis is the LOFAR Galaxy Zoo (LGZ) scale-up factor, which measures the total number of sources that the model selects for visual inspection divided by the number that we should really inspect. This is the combination of parameters that we aim to minimise. Symbols are as in Fig.~\ref{fig:roc_dr1}.}
  \label{fig:fdr_vs_LGZ}
\end{figure}

Instead of evaluating the whole performance of the model solely with the typical metrics (accuracy, precision, etc), we seek in particular to minimise the number of sources wrongly predicted as suitable to process with LR while keeping the number of sources sent to visual inspection low. These two requirements can be captured by: (i) the False Discovery Rate, FDR = FP/(FP+TP), which quantifies the fraction of sources sent to LR which are incorrect; and (ii) a parameter we refer to as the LOFAR Galaxy Zoo scale-up factor, given by (TN+FN)/(TN+FP), which expresses the factor by which the number of sources selected for visual analysis in LGZ is higher than the number actually required to be sent (cf. Table~\ref{tab:FC_groups}). In Fig.~\ref{fig:fdr_vs_LGZ} we show how the comparison between these two metrics changes as we change the cut-off threshold (open symbols, colour-coded by threshold level).

Although Fig.~\ref{fig:fdr_vs_LGZ} does not dictate which threshold value to use, the practical requirement to keep the LGZ scale-up factor to below $\sim 2$ pushes for a lower value of the threshold than the nominal 0.5 value, while the threshold values should not be so low to allow a false discovery rate above about 1 per cent. In practice, we adopt a threshold value based on comparison with the \citetalias{williams2019lofar} decision tree results. After correction for recovered components (Sec.~\ref{sec:corrections}), the classifier outperforms the \citetalias{williams2019lofar} decision tree in {\it both} false discovery rate and LGZ scale-up factor for thresholds in the range 0.18 to 0.25. We select a threshold level of 0.20, as a round number towards the centre of this range.
This threshold value corresponds to an LGZ scale-up factor of 1.68 and a False Discovery Rate of 0.006 for the raw model outputs.

\subsection{Corrections adopted}
\label{sec:corrections}

Corrections were determined to account for the multi-component sources wrongly classified as suitable to cross-match by LR (FP) that would subsequently be recovered by LGZ. Specifically, we analyse the prediction for each \texttt{PyBDSF} component that makes up a multi-component radio source and if at least one of the components is sent to visual analysis by the model, the source is removed from the FP group. The sources recovered in this way are discussed in Sec.~\ref{sec:FP}; in many cases these are the cores of radio sources (which on their own resemble a compact radio source) for which the more extended lobes are sent to LGZ.

\begin{figure}
    \centering
    \includegraphics[width =0.6 \columnwidth]{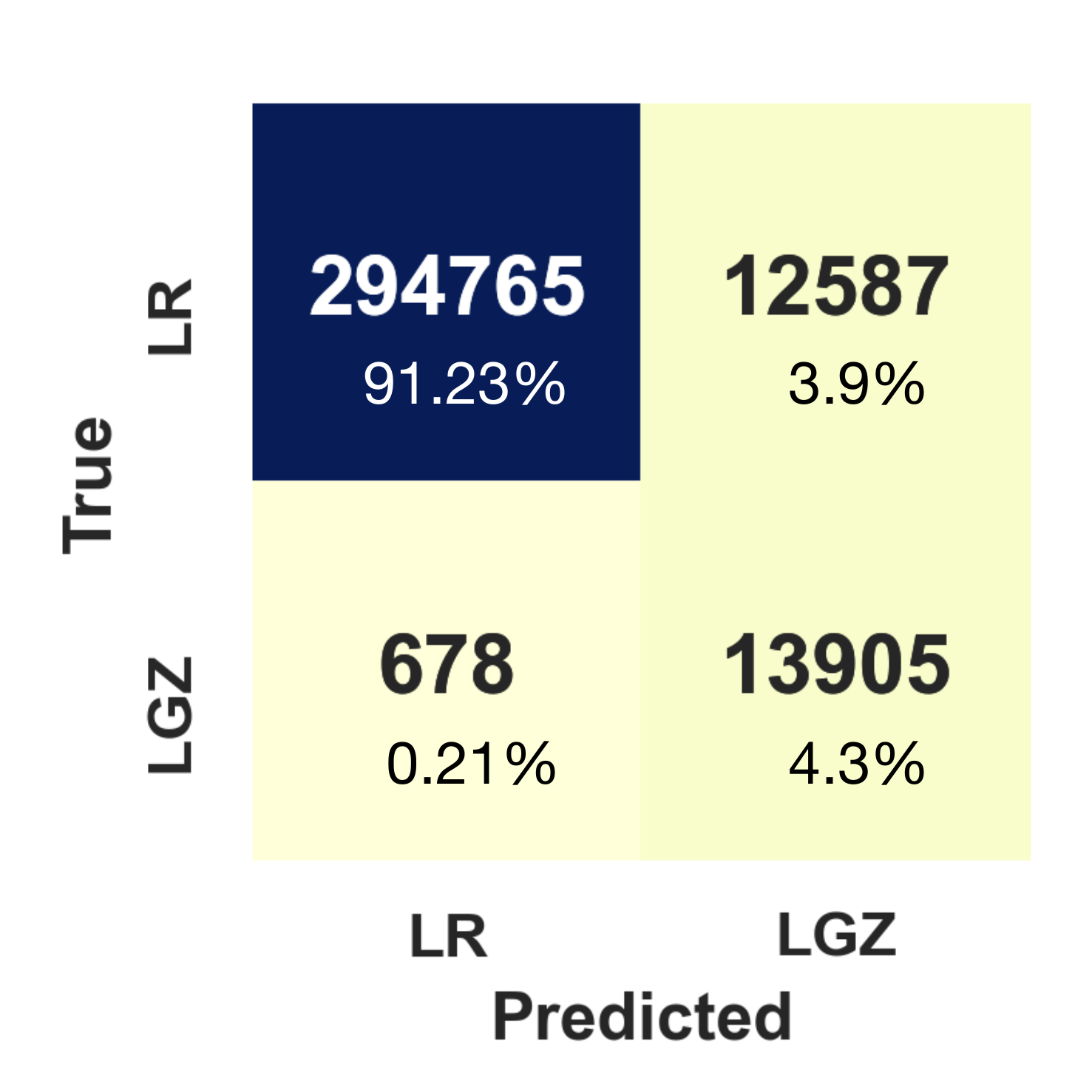}
    \caption{Confusion matrix for all the sources in LoTSS DR1 using the optimised model and a threshold value of 0.20 . The confusion matrix shows how examples belonging to each class are assigned correctly and incorrectly to the 2 possible classes. A perfect classifier would produce a confusion matrix filled diagonally with only TP (top left) and TN (bottom right) values, where the FP (bottom left) and FN (top right) would have values of zero, as defined in Sec.~\ref{sec:performance}. The background colours illustrates the proportion of sources in the matrix (given also by the percentage values in brackets) with darker colours representing a greater number of sources. The numbers presented correspond to the corrected values (see Sec.~\ref{sec:corrections}). }
    \label{fig:cm_20}
\end{figure}

We calculated the number of recovered sources for each different threshold value. The filled symbols on Fig.~\ref{fig:roc_dr1} demonstrate the improvement that these corrections make to the ROC curve analysis, and those on Fig.~\ref{fig:fdr_vs_LGZ} demonstrate the impact on our metric plot (FDR vs  LGZ scale-up factor) after applying these corrections to the FDR. Except at the very lowest thresholds, the improvement that the corrections make to the FDR is very significant; the fraction of recovered sources increases for higher threshold values, leading to very low FDRs at high thresholds, but with the cost of a higher number of visual inspections. For our adopted threshold of 0.20 , 63 per cent of the FP sources are recovered, resulting in a corrected false discovery rate of 0.002. This is also shown in the confusion matrix for that cut-off level, presented in Fig.~\ref{fig:cm_20}. In the analysis that follows, these corrections are applied unless stated otherwise.

\subsection{Performance relative to W19 decision tree}
\label{sec:flowchart_comparison}

\begin{figure}
    \centering
    \includegraphics[width = 1.0 \columnwidth]{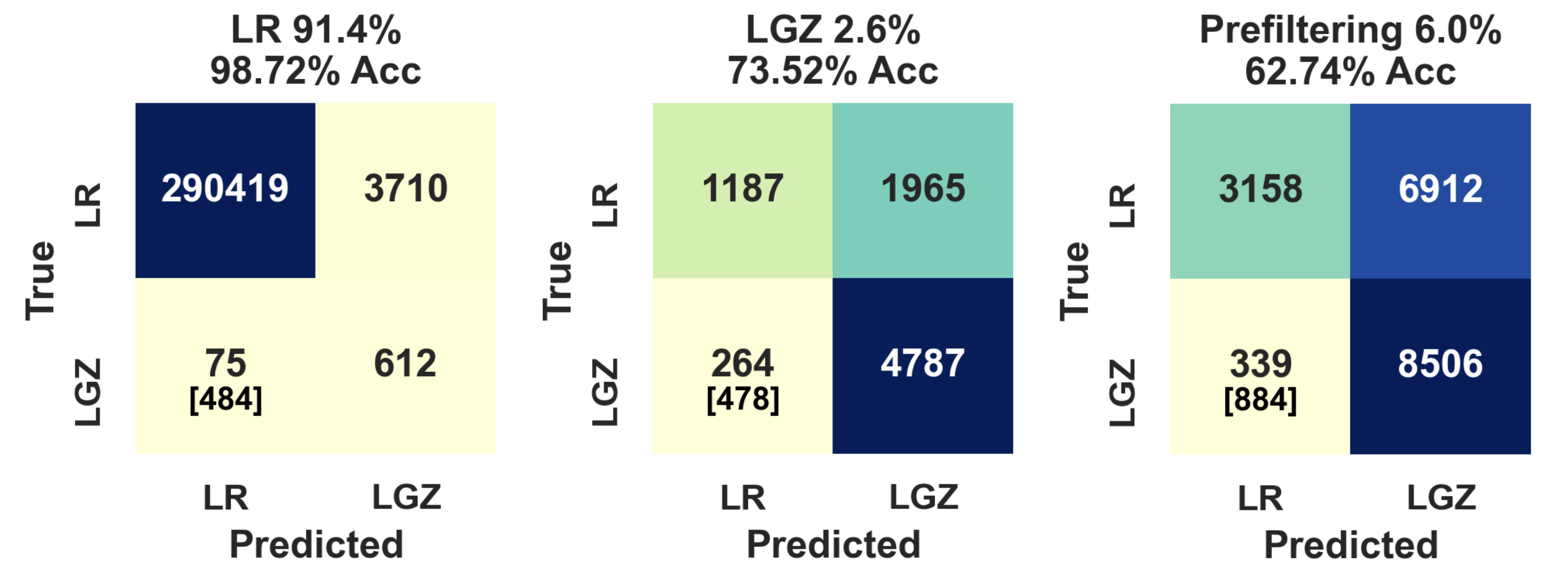}
    \caption{The model confusion matrix (for a threshold level of 0.20 ), split by the three main decision tree outcomes of \citetalias{williams2019lofar}: LR, LGZ and prefiltering. The FP values quoted are after corrections, with the numbers in brackets showing the values prior to corrections. As may be expected, the highest classification accuracy is for the LR sources, and the lowest accuracy is for the population of sources with intermediate parameter values deemed by \citetalias{williams2019lofar} to require prefiltering.} 
    \label{fig:cm_fc_groups}
\end{figure}

In this section we compare the performance of our model against that of the \citetalias{williams2019lofar} decision tree for the same dataset. First, in Fig.~\ref{fig:cm_fc_groups} we present the confusion matrix for the final model, split by the three main decision tree outcomes of \citetalias{williams2019lofar}: suitable for LR, send to LGZ, or requires prefiltering. 

It can be seen that the performance of the model on the `LR group' is excellent with nearly 99 per cent of the sources being deemed by the classifier to be suitable for LR. Furthermore, of the 1,096 sources that were incorrectly selected by the \citetalias{williams2019lofar} decision tree as `LR' but which were subsequently re-classified during the LGZ process (e.g. by being examined in parallel with another LGZ source) the classifier correctly sends the majority (over 600 sources) to LGZ, and of the rest all but 75 are recovered by having an alternate component of the source sent to LGZ. The classifier does send 3,710 sources to LGZ that \citetalias{williams2019lofar} sent directly to LR and which have a label of being suitable for LR. However, it is important to note that none of these sources has been visually examined to confirm that the \citetalias{williams2019lofar} label is correct: where the \citetalias{williams2019lofar} decision tree provided a LR classification, that was simply adopted by \citetalias{williams2019lofar} (unless LGZ examination of a different \texttt{PyBDSF} component over-rode that). There may, therefore, be (many) examples amongst these 3,710 sources that, like the 1,096 sources discussed above, would have been re-labelled had they been visually examined and for which the classifier is therefore correct. We explore this further below, and in Sec.~\ref{sec:FP}.

For the sources selected by \citetalias{williams2019lofar} to go directly to the LGZ process, the classifier provides an overall accuracy of 73.5 per cent, with the lower value mostly driven by nearly 2,000 sources being sent to LGZ despite being suitable for LR. Nevertheless, amongst the 3,000 sources in the \citetalias{williams2019lofar} LGZ sub-sample that were found (after visual examination) to be suitable for LR, the classifier is able to send over one-third of these directly to LR, thus reducing the LGZ scale-up factor.

The classifier performance is poorest on the sources sent by \citetalias{williams2019lofar} for prefiltering. This is not surprising, since these are generally sources with intermediate parameter values, between the compact LR sources and the extended LGZ examples. Again the classifier is able to send around one-third of the true LR sources directly to LR, but still assigns nearly 7,000 sources incorrectly to the LGZ class, providing the largest contribution to the LGZ scale-up. The prefiltering category also contains the largest number of false positives (339 after corrections).

We also compare the performance of our model in our metrics of FDR vs LGZ scale-up factor, against those of the \citetalias{williams2019lofar} decision tree. The LGZ scale-up factor of the \citetalias{williams2019lofar} decision tree is easily calculated from the numbers in Table~\ref{tab:FC_groups} and corresponds to a value of 1.77, while the 1,096 \texttt{PyBDSF} sources identified as false positives implies a \citetalias{williams2019lofar} FDR of 0.004. The FDR and LGZ scale-up factors thus determined for the \citetalias{williams2019lofar} decision tree are shown in Fig.~\ref{fig:fdr_vs_LGZ}. Compared to these, the ML model with a threshold of 0.20  achieves both a lower false discovery rate and a lower LGZ scale-up factor. Furthermore, as discussed above, the value of 0.004 represents a lower limit to the FDR of \citetalias{williams2019lofar} because the objects selected as being suitable for LR analysis were, in general, not visually examined, and thus false positives were not identified. We can estimate the total number of FPs by assuming that the fraction of sources rescued in this way is broadly the same as the ML model (the `corrections' calculated in Sec~\ref{sec:corrections}). This value depends weakly on the threshold value adopted (for similar theshold values). For thresholds around 0.20  we calculated above that 63 per cent of the sources are rescued. If 1,096 sources correspond to 63 per cent then we can estimate the total number of false positives in the \citetalias{williams2019lofar} decision tree will be approximately 1,730 sources\footnote{Note that these extra false positives will be mis-labelled in the input dataset, most likely comprising some of the false negatives in the LR subset of Fig.~\ref{fig:cm_fc_groups} as discussed above, and thus the performance of the ML model may therefore be fractionally higher than quoted.}. This would correspond to a higher FDR of 0.006. 

\begin{figure*}
    \centering
    \includegraphics[width = 1 \textwidth]{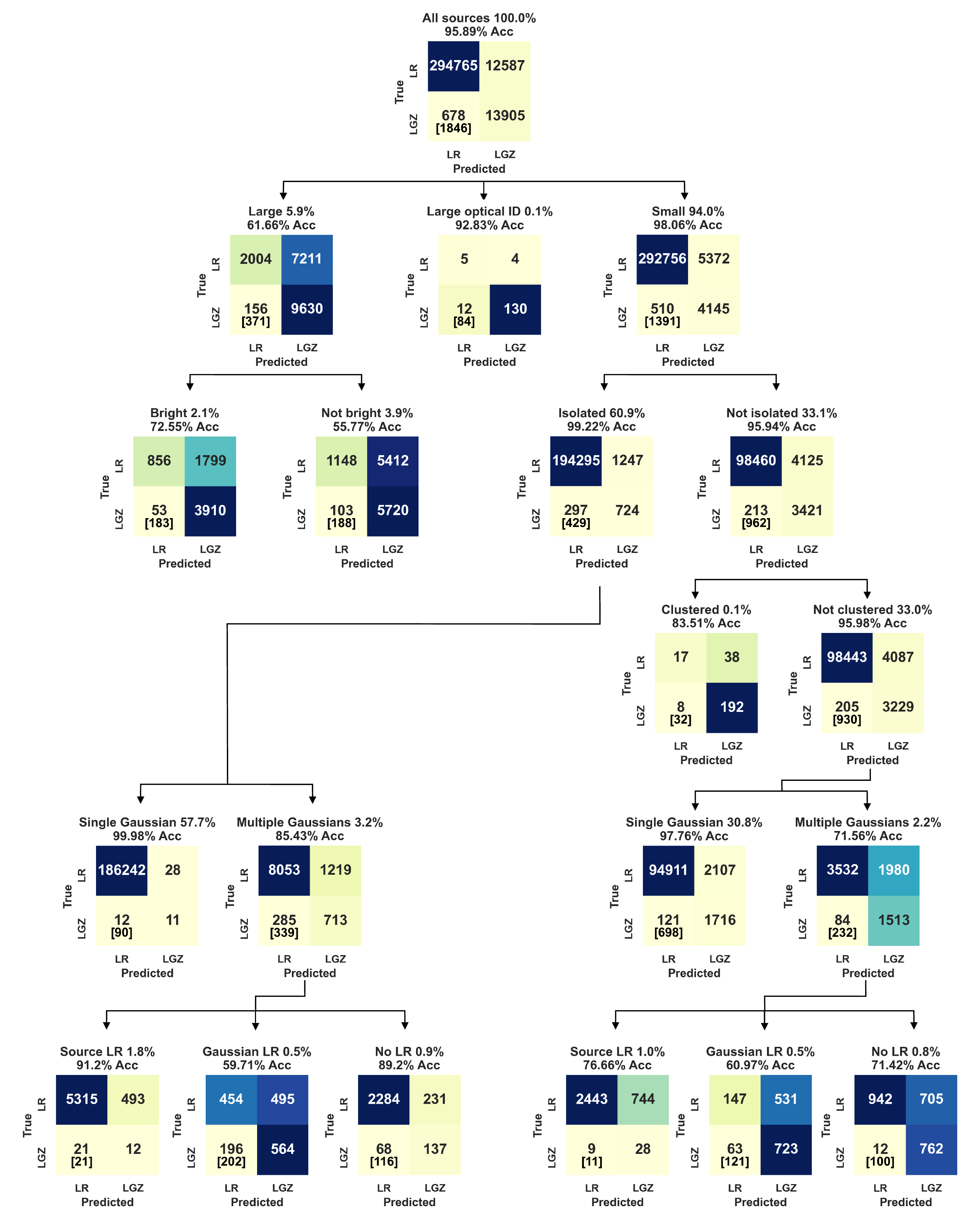}
    \caption{A simplified version of the \citetalias{williams2019lofar} decision tree, showing the performance of the classifier (in the form of the confusion matrix) at different locations on the decision tree.}
    \label{fig:CM_FC}
\end{figure*}

To gain a better understanding of which types of sources the model performs well on, and on which it performs badly, in Fig.~\ref{fig:CM_FC} we reproduce a simplified version of the \citetalias{williams2019lofar} decision tree, and examine the model confusion matrix at different locations of the decision tree. In the \citetalias{williams2019lofar} decision tree, sources are first classified as `Large' (major axis larger than 15 arcsec) or `Small' (under 15 arcsec), with a small number being associated with nearby large optical galaxies \citep[above 60 arcsec of radius in the 2$\mu$m all sky-survey extended source catalogue,][]{Jarrett2000}. The `Large' sources are then separated by \citetalias{williams2019lofar} in flux density (above or below 10\,mJy total flux densities), where \citetalias{williams2019lofar} send the brighter large sources all to LGZ and the fainter large sources all to prefiltering. The performance of the classifier on these two sub-categories is comparable to that on the general `LGZ' and `prefiltering' classes discussed above: these large sources produce more than half of the false negatives that lead to the above-unity LGZ scale-up factor. We examine the nature of these extended sources in more detail in Sec.~\ref{sec:perform_subsets}.

For the `Small' sources, \citetalias{williams2019lofar} next examined whether the source is relatively `Isolated' (no NN within 45 arcsec) or not. Isolated sources were examined to see if they were composed of single or multiple Gaussians. `Single Gaussians' were sent by \citetalias{williams2019lofar} to LR and it can be seen that the classifier achieves a remarkable accuracy of 99.98 per cent on these sources, which comprise nearly 58 per cent of the full sample. This subset of sources probably explains why the addition of LR features was found to only offer a small improvement in model performance in Sec.~\ref{sec:experiments}: these small, isolated, single Gaussian sources can almost entirely be sent for LR analysis based on their radio properties alone, and the LR provides no extra information. This does imply, however, that the addition of the LR features has much more impact in the other branches of the decision tree than the raw statistics of Table~\ref{tab:experiments} suggest -- indeed, for the `Large' and for multiple Gaussian sources, the addition of the LR information provides around 5 per cent increase in accuracy compared to the baseline.

For sources with multiple Gaussians, the \citetalias{williams2019lofar} decision tree was complicated, but can be simplified to consider those sources for which the \texttt{PyBDSF} source has a LR match above the LR threshold, those for which the \texttt{PyBDSF} source does not but one of the Gaussian components does, and those for which neither source nor any of the Gaussians has a LR match above the threshold. The classifier performs fairly well (accuracy $\approx 90$ per cent) on the first and third of these classes, but less well (accuracy $\approx 60$ per cent) on the Gaussian LR matches, which are only 0.5 per cent of the complete sample but contribute nearly 30 per cent of the (corrected) false positives in the whole sample. This implies that it may be possible to improve the classifier through better consideration of which Gaussian features to include (for example a second Gaussian to assist in identifying blended sources; see Sec.~\ref{sec:FP}) but such an investigation is beyond the scope of this paper. 

If the NN is within 45 arcsec (`Not Isolated') then the number of other \texttt{PyBDSF} sources within 45 arcsec is counted: sources with at least 4 others within that distance (`Clustered') were sent by \citetalias{williams2019lofar} to LGZ, and the classifier similarly sent most of these to LGZ. The `Unclustered' sources were then examined as for the isolated sources, into single or multiple Gaussian components and looking at the LR matches for the latter. In this case the performance on the single Gaussians (30 per cent of the overall sample) is less strong than for the isolated single-Gaussian sources, both in terms of false positives and LGZ scale-up, but still achieves 97.8 per cent accuracy. This illustrates that the near-neighbour components are impacting the classifier. Similarly, the performance on the multiple Gaussians is poorer than for the isolated sources (overall 71.6 per cent accuracy), in the sense of having a higher LGZ scale-up (more false negatives), albeit with a lower false positive rate.

\subsection{Performance as a function of source properties}
\label{sec:perform_subsets}

We also investigate the model performance as a function of source morphology and different source characteristics. For this, we consider the SOM, and separate the locations of the sources within this into six different morphological categories following \citet{mostert2021unveiling}. These six categories (described in more detail below) are: `extended singles'; `compact doubles'; `core-dominated doubles'; `large diffuse lobes'; `extended doubles'; and `single lobe / near neighbour'. Added on to these are the sources classified by \citet{Shimwell2019lofar} as `unresolved', which were not considered on the SOM.

\begin{figure}
    \centering
    \includegraphics[width =0.5 \columnwidth]{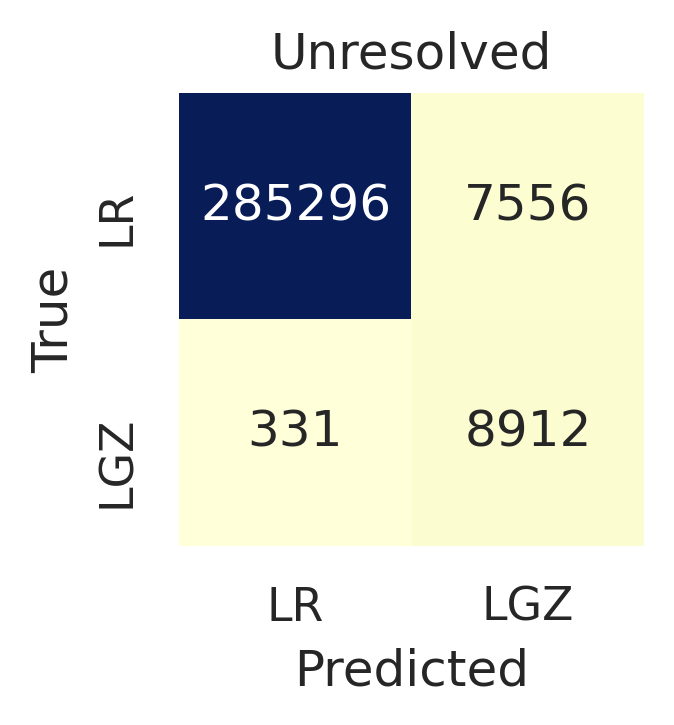}
        \includegraphics[width = 1 \columnwidth]{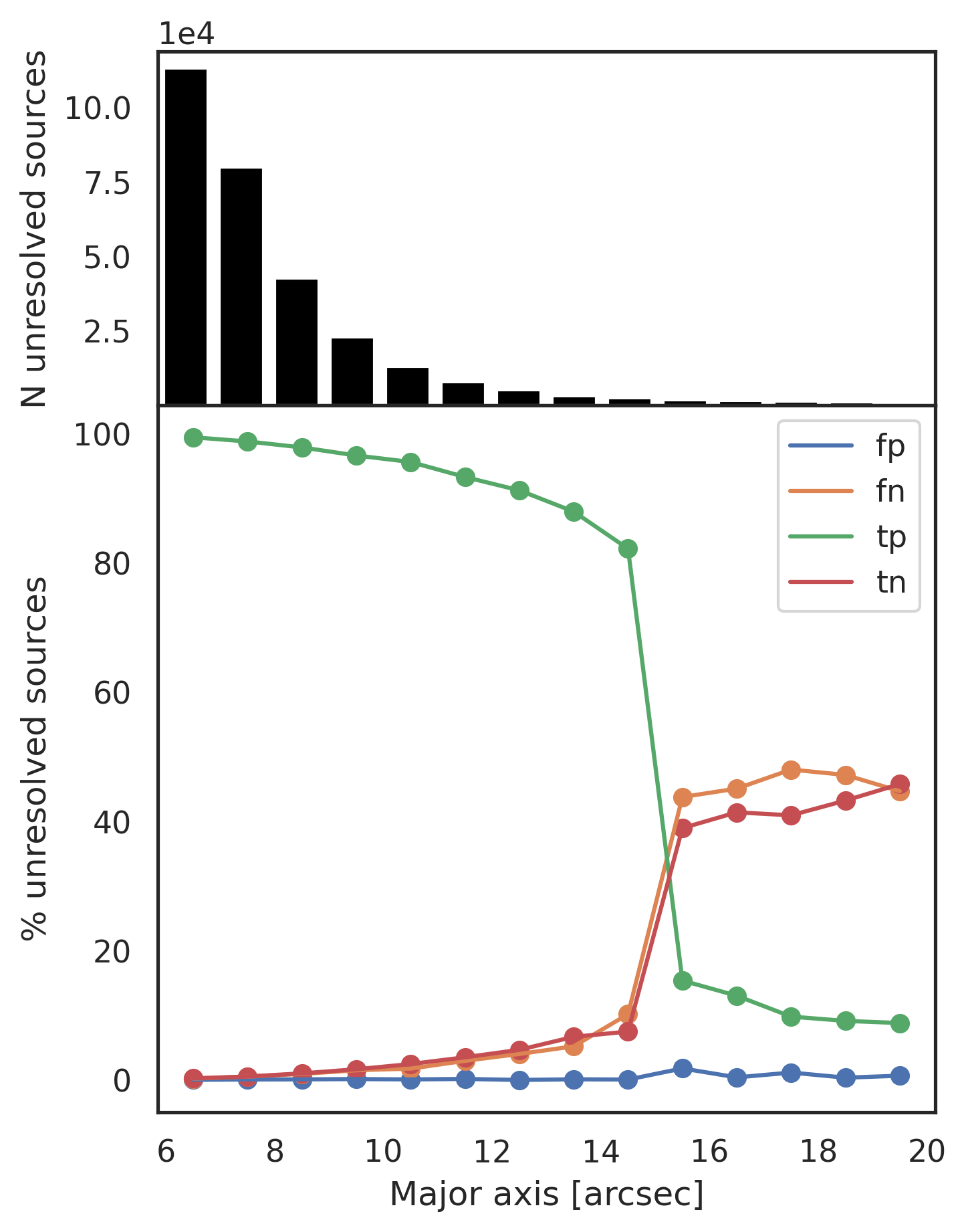}

    \caption{{\it Top:} the confusion matrix for the sources classified as `unresolved' by \citet{Shimwell2019lofar}. {\it Bottom:} the distribution of major axis sizes of these unresolved sources, and the variation of the different classifier outcomes as a function of the major axis size. The predicted LGZ outcomes are primarily associated with those sources with larger major axis sizes. The jump at a major axis of 15 arcsec is associated with the training sample characteristics; see text for more details. 
    Note that the $\approx$ 6,000 sources larger than 20 arcsec are not included on this plot.
    }
    \label{fig:cm_unresolved_som}
\end{figure}

Considering first the unresolved sources, the top panel of Fig.~\ref{fig:cm_unresolved_som} shows the confusion matrix for these sources. Perhaps surprisingly, more than 9,000 of these sources have `LGZ' labels, and the classifier also sends a further 7,556 sources to LGZ, corresponding to a significant proportion of the LGZ scale-up factor. To investigate the reason for this, in the bottom panel of Fig.~\ref{fig:cm_unresolved_som} we show how the different classifier outcomes vary with the size of the source major axis. Despite these sources being identified to be `unresolved' by \citet{Shimwell2019lofar}, the major axis sizes can extend to more than 20 arcsec; this is because the Shimwell et~al.\ classification adopts a signal-to-noise dependent size envelope for separating unresolved from extended sources based on their integrated flux density to peak brightness ratios, and so at low signal-to-noise where there is substantial scatter in the flux ratio it is possible to have quite large `unresolved' sources. It is not surprising that LR is not appropriate for these, as the radio position is poorly defined. Fig.~\ref{fig:cm_unresolved_som} indeed shows that both the true negative and false negative percentages increase with increasing major axis size, each reaching $\approx 10$ per cent at a major axis size of 15 arcsec. 

Fig.~\ref{fig:cm_unresolved_som} also illustrates that beyond 15 arcsec in size, the true negative fractions suddenly jump to 40 per cent. This is due to a feature of the training sample: all sources larger than 15 arcsec in size were visually examined by \citetalias{williams2019lofar}, and thus we expect them to be all correctly labelled, but at smaller sizes those sources for which the \citetalias{williams2019lofar} decision tree predicted `LR' were not visually examined; as discussed in Sec.~\ref{sec:threshold} some of these may be wrongly labelled. This suggests that the LGZ-fraction at sizes just below 15 arcsec may be somewhat higher than the labels suggest. 

Note that although the jump appears pronounced, only a small fraction of the `unresolved' sources have these large sizes, as can be seen in the histogram in Fig.~\ref{fig:cm_unresolved_som}.
Specifically, 10,516 (3.5 per cent  of the unresolved sources) have sizes between 12 and 15 arcsec, and 12,281 (4.1 per cent of the unresolved sources) are larger than 15 arcsec; these small numbers will not have a large effect on the classification outcomes. 
It is interesting to note that the false negative fraction shows a large jump at 15 arcsec size as well: due to the issues of the training sample, the classifier is learning that 15 arcsec is a critical size above which sources are more likely to require LGZ. This suggests that with an improved training sample that did not contain this issue, the performance of the classifier could potentially be improved even further than that presented here.

\begin{figure*}
    \centering
    \includegraphics[width = 1 \textwidth]{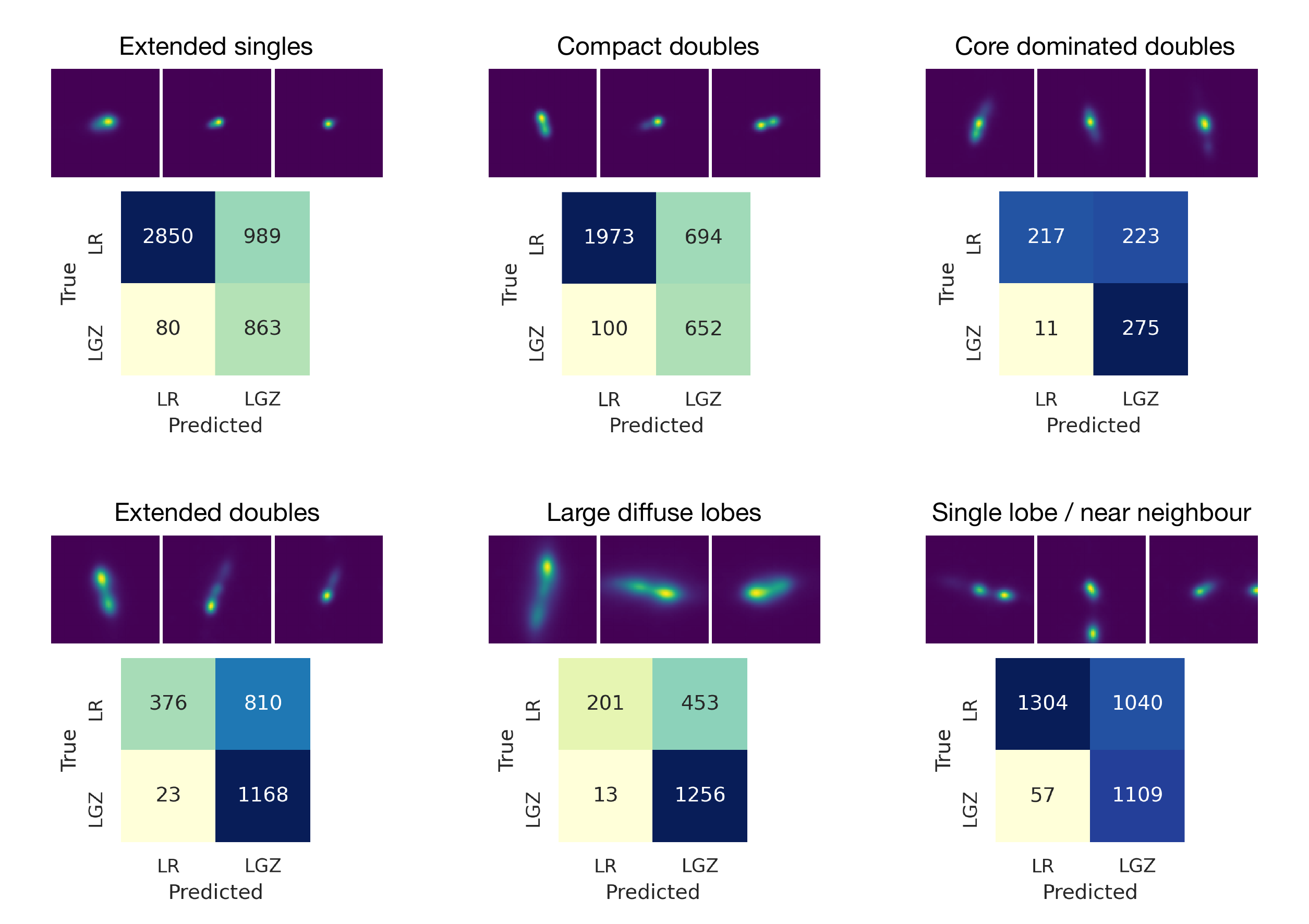}
    \caption{For six different broad morphological classes of extended sources defined by \citet{mostert2021unveiling}, the figure shows the confusion matrix, along with three example thumbnails drawn from the SOM representative sources.}
    \label{fig:cm_som}
\end{figure*}

Considering the extended sources, Fig.~\ref{fig:cm_som} displays the confusion matrices for each of the six categories of extended sources, along with three example thumbnails of each category, drawn from the SOM representative sources. For both the `extended single' sources (those fitted by \texttt{PyBDSF} as a single Gaussian, but classified by \citet{Shimwell2019lofar} as resolved) and the `compact doubles' (typically two Gaussian components in the \texttt{PyBDSF} source, but small angular size), the performance of the classifier is similar: over 75 per cent of both categories are classifiable by LR, and the classifier performs reasonably well (accuracy $\approx 77$ per cent) but sends about twice as many sources to LGZ as required. For the `core-dominated doubles', which show a bright central component but extended emission, the classifier sends about 70 per cent of the sources to LGZ, presumably due to the extended emission, although in reality 60 per cent would be classifiable by LR due to the central component (the other 40 per cent are not, as most are split into multiple \texttt{PyBDSF} sources). Similarly for the more `extended doubles', the classifier sends the majority to LGZ even though around half are symmetric enough that LR could be used. For the sources called `large diffuse lobes' by Mostert et~al.\ (which typically comprise either one or two extended lobes), the classifier achieves an accuracy of over 75 per cent by correctly sending the majority of the sources to LGZ, and again erring on the side of caution with an above-unity LGZ scale-up factor but few false positives. Finally. Mostert et~al.\ define a category of `single lobes', but we re-define this as `single lobe / near neighbour' because investigation reveals that while some of these are indeed one lobe of a double, two-thirds are single-component sources (classifiable by LR) for which there just happens to be a near neighbour. The classifier achieves a good accuracy (69 per cent) on these sources but again sends nearly twice as many as necessary to LGZ in order to minimise the number of false positives. Overall, it is clear that the performance on the extended sources is poorer than that on the `unresolved sources', but still relatively strong: the total LGZ scale-up factor for these extended sources is only $\approx 1.8$, not much higher than that of the unresolved sources, and the extended sources provide less than 300 false positives after corrections, with a false discovery rate below 4 per cent.

\subsection{Examination of False Positives}
\label{sec:FP}

\begin{figure*}
    \centering
    \includegraphics[width = 0.9 \textwidth]{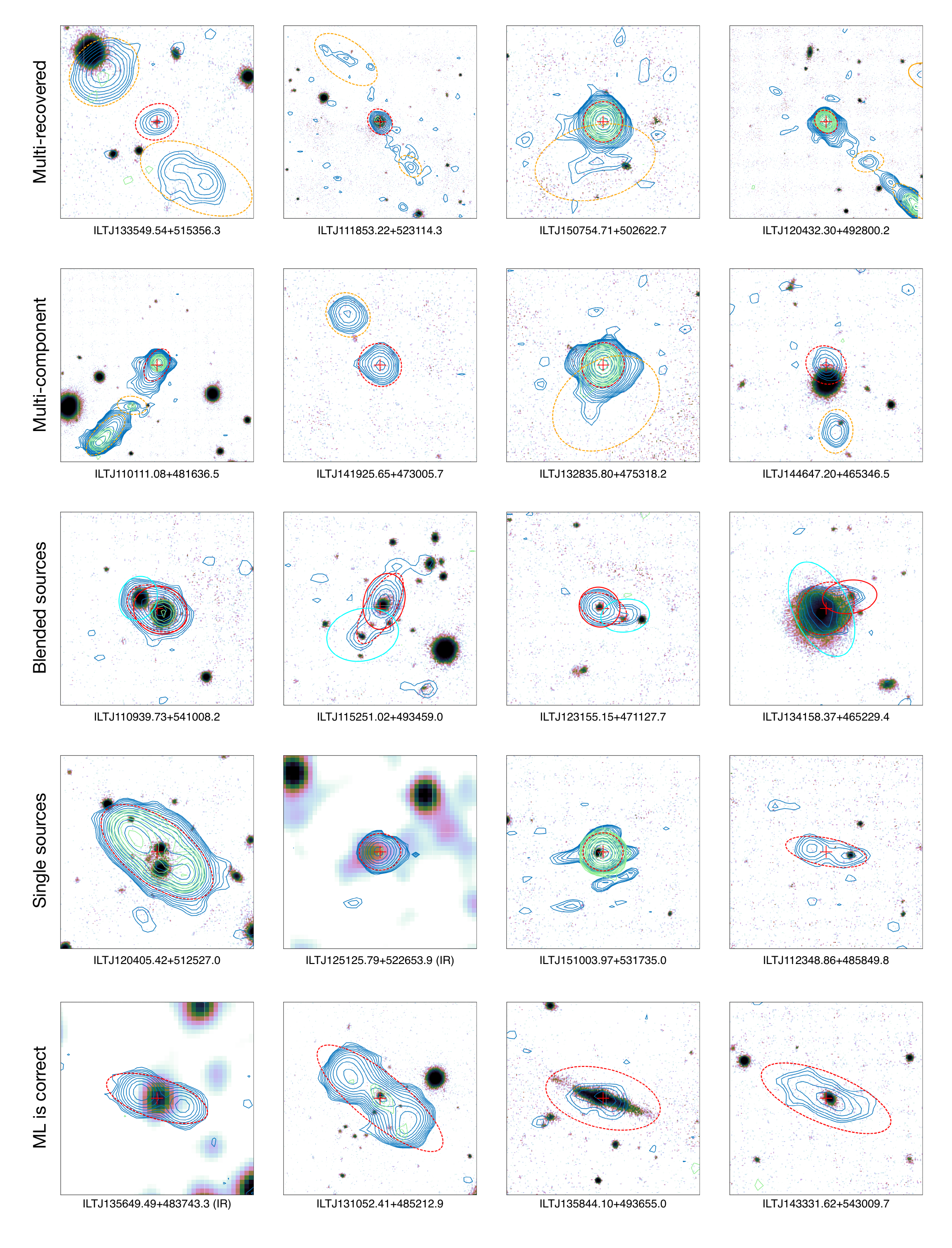}
    \caption{Examples of `false positive' classifications, where the model predicts that a LR approach is suitable, but in reality LR gives the wrong outcome and examination by LGZ is required. In all figures, the red cross and red dashed ellipse indicate the \texttt{PyBDSF} source being examined, the dark blue contours indicate the LOFAR radio emission, and the green contours indicate the higher frequency 1.4\,GHz radio emission from the FIRST survey. Yellow dashed ellipses indicate other \texttt{PyBDSF} sources that need to be combined to form a multi-component source; solid Yellow ellipses indicated unrelated sources. For the blended sources (row 3) the blue and red solid ellipses indicate the deblended components. The top row shows examples of multi-component sources where the false positive \texttt{PyBDSF} source is recovered (corrected) because a different component of the same source is sent to LGZ. The second row shows multi-component sources where none of the components is sent to LGZ. The third row shows blended sources, where LGZ is required to separate the \texttt{PyBDSF} source into two physical sources. The fourth row shows single components (correctly associated) but for which the LR prediction does not match the final \citetalias{williams2019lofar} ID outcome. For some of these, as indicated in the final row, the \citetalias{williams2019lofar} label appears to be incorrect and the machine learning (ML) correct. See text for further discussion.}
    \label{fig:FP}
    \vspace{-20pt}
\end{figure*}

Finally, in Fig.~\ref{fig:FP} we provide a montage of examples of false positive sources: these are the most critical failures, because of the lack of visual inspection. The false positives can be categorised into four main categories, illustrated in the first four rows of the figure. The top row of the figure shows examples of multi-component sources that get recovered (corrected) because one of the other \texttt{PyBDSF} components that makes up the source is sent to LGZ. These sources account for 63 per cent of all false positives. They are dominated by cases of the cores of radio sources for which the more extended lobes are sent to LGZ (see examples in the first and second columns), but also include sources showing small extensions selected as a separate \texttt{PyBDSF} source (third column; in some cases these may be noise and in other cases they may be genuine extensions), and even a small number of radio source lobes rescued by other components of the source (fourth column).

The second row shows additional multi-component sources, which are not recovered. In these cases, which amount to about 10 per cent of all false positives, it is essential to examine the sources with LGZ in order to properly associate the different \texttt{PyBDSF} sources into the same physical source and to identify the host galaxy, but the classifier predicts that all of the \texttt{PyBDSF} components are suitable for LR. These sources are typically relatively compact, two-component sources; sometimes it is clear from the radio structure that these form a single source (e.g. the example in the first column), whereas in many cases this is only apparent when examining the optical and infrared data and noting the presence of a host galaxy between the two lobes (examples in second and fourth columns). Finally, a proportion of these multi-component sources represent sources with weak extensions, some of which may be calibration artefacts (see example in the third column). In future work it would be worth investigating whether the performance of the classifier on these multi-component sources could be improved by including an additional feature related to the LR at the flux-weighted position between a source and its NN (corresponding to roughly where a host galaxy would be expected if the two sources form part of a double source).

The third row of Fig.~\ref{fig:FP} shows examples of blended sources (about 10 per cent of the false positives). These are cases where two physical sources have been merged into the same \texttt{PyBDSF} component, and these need to be examined and separated, but the classifier predicts that LR is appropriate. The optical images make the deblending requirement obvious, but it is understandable that this is difficult for the classifier to identify where the central component is substantially brighter in the radio and has a strong LR match. It is possible that if the LR of the second brightest Gaussian component was included as an additional feature of the classifier the performance on these objects could be improved.

The last two bottom rows represent sources that amount for about the remaining 20 per cent of the false positives. 
The fourth row presents examples of single sources (i.e. sources where \texttt{PyBDSF} has correctly identified the physical radio source) which the classifier predicts can be done by LR, but where the LR outcome disagrees with the final \citetalias{williams2019lofar} identification outcome. There can be many different reasons for this. The first column shows a source where the LR selects the more northerly galaxy, closer to the radio centroid, but examination of the radio contours led the LGZ participants to conclude that the southern galaxy is the true host. The second and third columns both give cases where the galaxy close to the radio centroid has a LR value above the threshold level, but the LGZ participants concluded that this was not sufficiently robust to accept, and found no ID. The fourth column shows an example where the LR finds no identification, but in LGZ it was concluded that this was an asymmetric source with the galaxy on the right-hand component being the host. It should also be noted that the LGZ process is not perfect and some of these single components may be mis-labelled, and should actually be true positives rather than false positives. The fifth row of Fig.~\ref{fig:FP} demonstrates this: these are all examples of single sources deemed by the classifier to be suitable for LR (and to have an identification) but judged by \citetalias{williams2019lofar} decision tree not to be. In all of these cases, the LR identification does appear to be robust. This suggests that these sources may be wrongly labelled by \citetalias{williams2019lofar}, and that the classifier is consequently performing even better than quoted. 

\subsection{Application to LoTSS DR2 subset}
\label{sec:application_to_DR2}

We have applied the model trained on LoTSS DR1 data directly to a subset of LoTSS DR2 in a small region in which LGZ source association and cross-matching process has already been completed for sources with total flux density higher than 4\,mJy \citep[these bright sources were examined first in LGZ in order to prepare targets for the WEAVE-LOFAR spectroscopic survey; ][]{smith2016weave}.
Since LoTSS DR2 contains almost 14 times more sources LoTSS DR1, the application of ML methods is crucial to help managing these large datasets.
We find that for the same threshold of 0.20, the classifier recommends that 11.8 per cent of LoTSS DR2 sources require visual analysis, compared to 8.2 per cent for LoTSS DR1. Investigation reveals that this difference is largely due to the source declinations: at declinations above about 50 degrees the classifier sends 9.7 per cent of LoTSS DR2 sources for visual analysis, which is not much higher than the DR1 statistics, but that fraction increases as we move to lower declinations. This declination dependence is likely to be largely due to the lower sensitivity of the LoTSS survey at lower declinations \citep{Shimwell2022}, which raises the median image rms. This means that a larger fraction of the detected sources are at higher flux densities where they are more likely to be multi-component and require LGZ (see upper-right panel of Fig.~\ref{fig:hist}). Adjusting the prediction threshold to a higher value would therefore help to increase the correct classifications at lower declinations. An additional factor may be the increasing size of the LOFAR beam at lower declinations (the use of deconvolved sizes in our features might have mitigated this).

To further test the performance of the model on DR2, we examine and compare the output predictions within this DR2 region. From a sample of 59,122 sources brighter than 4\,mJy, the classifier achieves an accuracy of 76 per cent; this compares with an accuracy of 82.7 per cent for sources brighter than 4\,mJy in DR1, without taking into consideration recovered source components in both cases. The lower accuracy for the DR2 data is mostly associated with the classifier sending more sources to LGZ, as discussed above. Considering that the classifier has not been trained on DR2, but simply applied with its DR1-determined hyperparameters (and the DR1 cut-off threshold) directly on the DR2 dataset, this shows that it has a strong ability to generalise to an unseen dataset. The optical cross-matching for LoTSS DR2 (Hardcastle et~al., in prep.) will differ from that of DR1 in the use of the DESI Legacy imaging surveys \citep{Dey_legacy2019} instead of Pan-STARRS as the primary optical survey; however, our use of `log$\_$lr$\_$tlv', that is, the logarithm of the ratio of the LR relative to the threshold value, as the primary LR feature should mitigate against these differences in the cross-match survey.

\section{Conclusions and future outlook}
\label{sec:conclusions}

In order to get the most science out of the survey catalogues being produced by the new generation of radio interferometers, it is necessary to properly associate radio source components into physical sources, and then cross-match those sources with multi-wavelength data. This enables us to identify the host galaxies and correctly derive the physical properties of the radio sources. To address the question of which sources are suitable for simple statistical cross-matching, and which ones require a more advanced (currently visual) approach, we trained a machine learning (ML) classifier using LoTSS DR1 and applied it to different LoTSS releases. The main conclusions of our work are:

\begin{itemize}
    \item Our best model is a tree-based gradient boosting classifier, and achieves an accuracy of 95 per cent on a balanced dataset. This accuracy is maximised by appropriate choice of features in the model: inclusion of information on nearest neighbour (NN) radio sources, on the properties of any LR match, and on the composition of the radio source in terms of Gaussian components all improve the model. 
    \item The full LoTSS dataset is highly imbalanced, with the majority ($\approx$95 per cent) of the sources being suitable for LR analysis. Adaption of the default 0.5 probability threshold for the classifier would result in far too many of these sources being predicted to require visual analysis. An optimised threshold of 0.20  restricts the LGZ sample to only 68 per cent larger than strictly required, while keeping the false discovery rate (i.e. the fraction of those sources accepted by LR that should have required LGZ) to only 0.2 per cent. With this threshold, the classifier outperforms the manually-defined decision tree used for LoTSS DR1 by \citetalias{williams2019lofar} in both the LGZ scale-up factor and the false discovery rate.
    \item We have investigated the performance of the classifier on sources of different radio morphologies and with different source characteristics. As expected, performance is strongest for the most compact sources, achieving an accuracy of over 98 per cent on sources with a major axis size smaller than 15 arcsec (and over 99.9 per cent on the subset of these that have no near neighbours and can be well-modelled by a single Gaussian). The accuracy drops to just above 60 per cent for sources larger than 15 arcsec in size, primarily due to sending substantially more sources to LGZ than required. 
\end{itemize}

The efficiency of the ML approach means that it can be applied to other radio surveys, and in particular to future data releases of the LoTSS survey, where the radio data are almost identical in nature to the DR1 sample analysed here \citep[although there will be small differences, associated with improvements in the calibration scheme and a changing telescope beam as we move to lower declination; see][for more details]{Shimwell2022}.
Because of these results, the classifier outcomes derived for the full DR2 sample have been used, in conjunction with the \citetalias{williams2019lofar} decision tree, to identify the LoTSS DR2 sources that are being sent to LGZ; Hardcastle et~al. (in prep.) will provide more details.

In conclusion, the ML classifier that we have developed has been shown to have a high accuracy at identifying those sources for which a statistical cross-match process is insufficient, and to outperform a manually-defined decision tree in both the false discovery rate, and in the number of sources that are predicted to require the time-consuming visual analysis step. The classifier has been demonstrated to be able to generalise to unseen datasets; it already has immediate application in the cross-matching of the LoTSS DR2 and can be easily applied to other radio surveys. 

The classifier could potentially be further improved by the inclusion of additional features, for example, the LR of a second Gaussian component to assist in identifying blended sources, a LR at the flux-weighted position between a source and its NN to help identify multi-component sources, or additional properties such as the local noise level or the source signal-to-noise ratio. However, even if the classifier were improved still further, the number of sources that require more than statistical cross-matching will still remain large, and visual analysis of all of these will become impractical as radio surveys continue to grow in size. The crucial next step is therefore to be able to replace visual analysis as the process to handle those sources. To this end, work to automatically associate multi-component sources
\citep[e.g.][]{Mostert2022} and to improve automatic source cross-matching for extended sources \cite[e.g. ridge-line based approaches;][]{barkus2022application} is on-going.
The automatic source association of \citet{Mostert2022} actually makes use of the ML classifier developed here to reject unassociated compact sources that lie within the boundary of more extended multi-component sources. It is likely that a selection of different ML and deep learning techniques will need to be developed and combined to fully solve this problem.



\section*{Acknowledgements}

We appreciate the valuable comments made by the anonymous reviewer.
LA is grateful for support from the UK Science and Technology Facilities Council (STFC) via CDT studentship grant ST/P006809/1.
PNB and JS are grateful for support from the UK STFC via grants ST/R000972/1 and ST/V000594/1.
WLW acknowledges support from the CAS-NWO programme for radio astronomy with project number 629.001.024, which is financed by the Netherlands Organisation for Scientific Research (NWO). 
MJH and DJBS acknowledge support from  the UK STFC under grant ST/V000624/1.
RK acknowledges support from the UK STFC via studentship grant ST/R504737/1.
This work made use of the Scikit-learn machine learning python library \citep{scikit-learn}; the Astropy python package for Astronomy \citep{astropy:2013, astropy:2018}; and the Pandas library for data manipulation and analysis \citep{pandas}. Plots were made with the help of matplotlib \citep{Matplotlib} and seaborn \citep{seaborn}.
LOFAR data products were provided by the LOFAR Surveys Key Science project (LSKSP; \hyperlink{https://lofar-surveys.org}{https://lofar-surveys.org}) and were derived from observations with the International LOFAR Telescope (ILT). LOFAR \citep{vanHaarlen2013lofar} is the Low Frequency Array designed and constructed by ASTRON. It has observing, data processing, and data storage facilities in several countries, that are owned by various parties (each with their own funding sources), and that are collectively operated by the ILT foundation under a joint scientific policy. The ILT resources have benefitted from the following recent major funding sources: CNRS-INSU, Observatoire de Paris and Universit{\'e} d'Orl{\'e}ans, France; BMBF, MIWF-NRW, MPG, Germany; Science Foundation Ireland (SFI), Department of Business, Enterprise and Innovation (DBEI), Ireland; NWO, The Netherlands; The Science and Technology Facilities Council, UK; Ministry of Science and Higher Education, Poland.
For the purpose of open access, the author has applied a Creative Commons Attribution (CC BY) licence to any Author Accepted Manuscript version arising from this submission.



\section*{Data availability}

The tabular data underlying this article are provided in the supplementary online material (see Table \ref{tab:master_table} for column description). The datasets were derived from LoTSS Data Release 1 publicly available at https://lofar-surveys.org/dr1$\_$release.html.



\vspace{-0.5cm}
\renewcommand*{\bibfont}{\footnotesize}
\setlength{\bibsep}{1pt}
\bibliographystyle{mnras} 
\bibliography{bibiography.bib}



\appendix
\section{Machine Learning tools and algorithms}
\label{app:ML}

\subsection{AutoML}

In Sec.~\ref{sec:experiments}, we streamline model selection and optimisation using Automated Machine Learning (AutoML). AutoML generates optimal ML pipelines by identifying the best model and model hyperparameters. AutoML has already been used in astronomy with the application of open source AutoML toolkits, and the use of AI platforms. For instance, \cite{Arsioli2020} investigated the \texttt{Ludwig} framework \citep[][]{molino2019ludwig} in the classification of blazars, and \cite{Zuntz2021} used \texttt{Auto-keras} \citep[][]{jin2019auto} to select one of the models for the LSST-DESC 3x2pt Tomography Optimization Challenge. \cite{tarsitano2022} used the \texttt{modulos.ai} platform to select the best CNN architecture to perform optical galaxy morphological classification, \cite{barsotti2022} used the \texttt{DataRobot} platform to predict gravitational waveforms from compact binaries, and \cite{kruk2022hubble} used the \texttt{Google Cloud AutoML Vision} to train a CNN for the Hubble Asteroid Hunter project. 
Other AutoML framework examples include the Tree-based Pipeline Optimization Tool \citep[\texttt{TPOT},][]{OlsonGECCO2016} and \texttt{Auto-sklearn} \citep[][]{feurer2015efficient} for traditional ML; and \texttt{Auto-Pytorch} \citep[][]{zimmer2021auto} for deep learning. In this work we use \texttt{TPOT}, which we will explain in more detail next.

\subsubsection{TPOT}

\texttt{TPOT} is an open source AutoML tool that evaluates different ML pipelines using genetic programming \citep[GP,][]{banzhaf1998genetic}. In the field of evolutionary computation, GP (and its variants) are the most widely used type of evolutionary algorithm \citep[e.g.][]{eiben2015evolutionary}. By using a function that minimises the error in the solution, these algorithms search for an optimal candidate within a group of potential solutions. 
\texttt{TPOT} was further developed to incorporate pipeline design automation; it performs feature selection, preprocessing and engineering, besides algorithm searching and optimisation. 
It uses the Python \texttt{Scikit-Learn} library to implement both individual and ensemble tree-based models (Decision Trees, Random Forests and Gradient Boosting), non-probabilistic and probabilistic linear models (Support Vector Machines and Logistic Regression), k-nearest neighbours; and it uses \texttt{PyTorch} for neural networks. The code can be used for both classification and regression problems, and has been adapted to work with large datasets of features \citep[][]{le2020scaling}. 

\texttt{TPOT} is built on Distributed Evolutionary Algorithms in Python \citep[DEAP,][]{de2012deap}, a framework that implements evolutionary computation. \texttt{TPOT} implements GP by creating trees of pipeline operators and evolving those operators in order to maximise accuracy. In brief \citep[see][for more details]{OlsonGECCO2016,Olson2016TPOT}, in the first iteration (i.e. generation) \texttt{TPOT} sets and evaluates a random number of tree-based pipelines (i.e. population). The next generation is constructed as follows. First, 10 per cent of the new pipelines are copies of the highest accuracy pipeline from the previous generation; 90 per cent of new pipelines are selected from the previous generation using a 3-way tournament selection with a 2-way parsimony (i.e. 3 random pipelines are evaluated by first eliminating the one with the poorest performance and then choosing the simplest of the remaining two). Next, a proportion of these new generation pipelines are modified; 5 per cent of the pipelines suffer a one-point \textit{crossover}, which consists of swapping the contents of two random pipelines at a random split in the tree of operators. For 90 per cent of the remaining new pipelines a \textit{mutation} is applied, where random operators are inserted, removed or replaced in the pipelines. The process is repeated for the number of generations defined. 

\subsection{Ensembles of decision trees}
We provide a brief description of ensembles of decision trees, with particular focus on the Gradient Boosting Classifier (GBC), which is the type of algorithm we chose to apply (see Sec.~\ref{sec:experiments}). Ensembles of decision trees are sets of decisions trees, typically containing between 100 and 1,000 trees. On their own, individual trees have moderate performance, but when combined, the ensemble achieves strong performance. There are different ways of creating these ensembles. Two common techniques are bagging and boosting. In bagging (e.g. Random Forest) the trees are created in parallel using splits between the features and the final prediction is, in general, given by the average of the predictions or the majority of the votes of the trees. By contrast, in boosting (e.g. Gradient Boosting) each tree is constructed sequentially by minimising a loss function from the preceding tree, and in general trees (also referred as \textit{week learners}) with better performance have higher weight on the final predictions. \citep[see e.g.][for details and comparison of the methods]{bauer1999empirical, sutton2005classification}. 
Since bagging models output average predictions, they reduce the variance of the model, and are therefore more robust to outliers and defective features (since these will be mainly ignored). In boosting, the trees grow in the direction where the loss is minimised. Therefore, each additional tree reduces the bias of the model. By aggregating the predictions from all the trees, boosting also reduces the model variance \citep[][]{schapire2013boosting}. As a consequence, boosting models are more powerful than bagging models, but they can also overfit in some cases, especially when the number of trees is increased: since each iteration reduces the training error, this can be made arbitrarily small by growing trees, which can lead to overfitting to the training data \citep[][]{trevor2009elements}.

The model used in this work is a GBC, for which the original formulation can be found in \cite{friedman2001greedy}. It is a stochastic boosting model \citep[][]{friedman2002stochastic} which uses a functional gradient descent \citep[][]{mason1999boosting}. Consider an input training set of $n$ examples, where each example has a set of feature values $x$ and an output value $y$ (where for our binary classifier $y$ is defined as 0 or 1). 
The model sequentially builds an ensemble of weak learners, whose output prediction after iteration $m$ is $F_m$.

The weak learners are constructed by first initialising a very simple model ($F_0$) in which the output prediction is a constant for all sources; this constant may be set to zero or may be chosen to minimise the initial  loss function $L_0$. The loss function is defined based on the difference between predicted and true values, summed across the full training population: for the binary classifier used in this work, a binary log loss function (also known as binary cross-entropy or binomial deviance) is used:

\begin{equation}
L_m = \frac{1}{n} \sum_{i=1}^{i=n} y_i \log F_m,i + (1-y_i) \log(1-F_{m_i})
\end{equation}

\noindent where $L_m$ is the loss function for tree $m$ and $F_{m_i}$ is the model prediction for source $i$ in iteration $m$. 
For each subsequent iteration, $m$, the procedure is then as follows. First, the {\it pseudo-residuals} for each training source are calculated from the model. Pseudo-residuals ($r$) for each source $i$ are defined as

\begin{equation}
r_{i,m} = \frac{\partial L(y_i, F_{m-1}(x_i))}{\partial F_{m-1}(x_i)}
\end{equation}

\noindent A modified dataset is then made with the input parameters $x$, and output values of $r$. A tree is then fitted to this dataset, with the resultant predictions $h_m(x_i)$. Using these predictions, the model prediction $F_m$ is defined as

\begin{equation}
F_m(x) = F_{m-1}(x) + \nu h_m(x)
\end{equation}

\noindent where $\nu$ is the shrinkage parameter, commonly referred as learning rate, which scales the contribution of each tree by a factor between 0 and 1, acting as a regularisation method \citep[][]{friedman2002stochastic}. This value must be such that there is a trade-off with the number of trees $M$ in the model.
The loss function for the new tree can then be calculated, and the process is repeated until a final prediction $F_M(x)$ is produced. Due to the way that the model is constructed, it can be considered to be a weighted additive combination of all of the individual weak learners from which it is comprised:

\begin{equation}
F_M(x_i)=\sum_{m=1}^M \nu h_m(x_i), 
\end{equation}

\section{Master table}

An electronic table provides the source identification and feature data used as input to the ML algorithm, along with the source identification flags and diagnostic flags, and the final model prediction. Table~\ref{tab:master_table} describes the columns provided in that table, which also include the columns from Table~\ref{tab:features}.

\begin{table*}
 \centering
 \caption{Master table columns description. These were selected or computed using different catalogues:
 a - LoTSS DR1 PyBDSF radio source catalogue \citep{Shimwell2019lofar}; 
 b - LoTSS DR1 PyBDSF Gaussian component catalogue \citep{Shimwell2019lofar};
 c - LoTSS DR1 Gaussian and PyBDSF LR catalogues \citepalias{williams2019lofar};
 d - Optical LoTSS DR1 source catalogue \citepalias{williams2019lofar};
 e - LoTSS DR1 self-organised map \protect\citep[SOM;][]{mostert2021unveiling};
 f - results calculated in this work. Note that artefacts are flagged with the value of -99.}

 \label{tab:master_table}
 \begin{tabular}{ll}
  \hline
  Column & Definition \& Origin\\
  \hline
  \textbf{Source information} & \\
  Source$\_$Name & PyBDSF source identifier (typically a combination of RA and DEC position)$^a$ \\
  RA & PyBDSF source right ascension [deg]$^a$  \\
  DEC & PyBDSF source declination [deg]$^a$   \\
  Source$\_$Name$\_$final & Final radio source name (after any source association or deblending); NULL if artefact $^d$ \\
  RA$\_$final & Final radio source right ascension [deg]$^d$  \\
  DEC$\_$final & Final radio source declination [deg]$^d$ \\
  AllWISE$\_$lr & Source identifier of near-infrared AllWISE counterpart cross-matched by likelihood ratio$^{c,f}$ \\
  AllWISE$\_$final &  Source identifier of finally-assigned near-infrared AllWISE counterpart$^d$\\
  objID$\_$lr & Source identifier of optical Pan-STARRS cross-match by likelihood ratio$^{c,f}$ \\
  objID$\_$final & Source identifier of finally-assigned optical Pan-STARRS counterpart$^d$ \\
  Mosaic$\_$ID & HETDEX mosaic which contains the source image$^d$\\
  Gaus$\_$id & Identifier of PyBDSF Gaussian component used as feature$^b$ \\
  NN$\_$Source$\_$Name & PyBDSF Source$\_$Name of the Nearest Neighbour$^a$\\
  \hline
  \textbf{Identification flags} & \\
W19dt & \citetalias{williams2019lofar} decision tree main outcomes [0-LGZ, 1-LR (ID or no ID), 2-prefiltering, 3-large optical IDs, -99-artefacts]$^d$\\
  \textbf{Diagnosis Flags} & \\
  association & PyBDSF source association diagnosis [1-single, 2-blended, 4-multi component, -99-artefacts]$^f$\\
  accept$\_$lr & Source suitable to LR technique [0-False,1-True, -99-artefact]$^f$\\
  multi$\_$component & Multiple component source [0-False,1-True, -99-artefact]$^f$ \\
  \hline
   \textbf{ML features} & \\
 (Several columns) & Machine learning features from Table~\ref{tab:features} \\
  \hline
  \textbf{Additional ML features} & \\
  n$\_$gauss & Number of Gaussians that compose the PyBDSF source$^b$ \\
  gauss$\_$total$\_$flux & Gaussian flux density of Gaussian component used as feature [mJy]$^b$\\
  
  \textbf{Deconvolved sizes} &  \\
  DC$\_$Maj & PyBDSF deconvolved major axis [arcsec]$^a$ \\
  DC$\_$Min & PyBDSF deconvolved minor axis [arcsec]$^a$\\
  gauss$\_$dc$\_$maj & Gaussian deconvolved major axis [arcsec]$^b$ \\
  gauss$\_$dc$\_$min & Gaussian deconvolved minor axis [arcsec]$^b$\\
  
  \textbf{Likelihood Ratio (LR) values}\\
  lr & LR value match for the PyBDSF source$^{c,f}$\\
  gauss$\_$lr & LR value match for the Gaussian$^{c,f}$\\
  highest$\_$lr & Highest LR value match between the Gaussian and the source$^{c,f}$\\
  NN$\_$lr & LR value match for the PyBDSF nearest neighbour$^{c,f}$\\
  \textbf{Self-Organising Map (SOM)} & \\
  10x10$\_$closest$\_$prototype$\_$x & Row position of the PyBDSF source on the LoTSS DR1 cyclic 10x10 SOM$^e$\\
  10x10$\_$closest$\_$prototype$\_$y & Column position of the PyBDSF source on the LoTSS DR1 cyclic 10x10 SOM$^e$\\
  \hline
  \textbf{Predictions} & \\
  probability$\_$lr & Prediction probabilities to accept the LR      match [range 0-1, 0-False and 1-True]$^f$ \\
  dataset & Dataset splitting [0 - not on the train or test sets, 1-training set, 2-test set]$^f$ \\
  prediction$\_$0.20 & Predictions for 20\% threshold [0-send to LGZ, 1-accept LR, 2-recovered component]$^f$ \\
  \hline
 \end{tabular}
\end{table*}


\bsp	
\label{lastpage}
\end{document}